\documentclass[journal,draftcls,onecolumn,12pt,twoside]{IEEEtranTCOM}
\normalsize
\usepackage{cite}
\usepackage{graphicx}
\usepackage{psfrag}
\usepackage{subfigure}
\usepackage{url}
\usepackage{stfloats}
\usepackage{amsfonts,amssymb,amsmath,bm,paralist,theorem,cite,ifthen,color}
\usepackage{array}
\usepackage{calc}
\usepackage{longtable}
\usepackage{stfloats}  
\usepackage{graphics,booktabs,color,epsfig,subfigure}
\usepackage{multirow}
\usepackage{algorithmic,algorithm}
\usepackage{setspace} 
\usepackage{subfig}
\usepackage{pst-pdf}
\usepackage{ifplatform}
\usepackage{xcolor}
\usepackage{preview}
\usepackage{environ}
\usepackage{graphicx}

\usepackage{caption}

\usepackage{epstopdf}

\newtheorem{lemma}{Lemma}
\newtheorem{proposition}{Proposition}
\newtheorem{Theorem}{Theorem}
\newtheorem{Definition}{Definition}
\newtheorem{Corollary}{Corollary}

\theorembodyfont{\rmfamily}

\newtheorem{defn}{Definition}

\newtheorem{remark}{Remark}

\newtheorem{problem}{Problem}

\newcommand{\bfF}{ \mathbf{F} }

\newcommand{\bfC}{ \mathbf{C} }

\newcommand{\beq}{\begin{equation}}
\newcommand{\eeq}{\end{equation}}

\makeatletter
\newenvironment{breakablealgorithm}
  {
   \begin{center}
     \refstepcounter{algorithm}
     \hrule height.8pt depth0pt \kern2pt
     \renewcommand{\caption}[2][\relax]{
       {\raggedright\textbf{\ALG@name~\thealgorithm} ##2\par}%
       \ifx\relax##1\relax 
         \addcontentsline{loa}{algorithm}{\protect\numberline{\thealgorithm}##2}%
       \else 
         \addcontentsline{loa}{algorithm}{\protect\numberline{\thealgorithm}##1}%
       \fi
       \kern2pt\hrule\kern2pt
     }
  }{
     \kern2pt\hrule\relax
   \end{center}
  }
\makeatother

 \usepackage{tikz}
\usetikzlibrary{arrows}
\usetikzlibrary{chains}
\usetikzlibrary{positioning,datavisualization}
\usetikzlibrary{calc}
\usetikzlibrary{fadings,shapes.arrows,shadows} 
\usetikzlibrary{shapes}
\usepackage{ellipsis}
\usetikzlibrary{calc}
\usetikzlibrary{decorations.pathreplacing,decorations.markings,shapes.geometric}

\tikzset{naming/.style={align=center,font=\small}}
\tikzset{antenna/.style={insert path={-- coordinate (ant#1) ++(0,0.25) -- +(135:0.25) + (0,0) -- +(45:0.25)}}}
\tikzset{station/.style={naming,draw,shape=dart,shape border rotate=90, minimum width=10mm, minimum height=10mm,outer sep=0pt,inner sep=3pt}}
\tikzset{mobile/.style={naming,draw,shape=rectangle,minimum width=5mm,minimum height=5mm, outer sep=0pt,inner sep=3pt}}
\tikzset{radiation/.style={{decorate,decoration={expanding waves,angle=40,segment length=1pt}}}}

\newcommand{\UE}[1]{%
\begin{tikzpicture}[every node/.append style={rectangle, minimum width=0 pt}]
\node[mobile,scale=0.3] (box) {#1};

\draw [scale=0.25](box.north east) [antenna=1];
\end{tikzpicture}
}

\newcommand{\BS}[1]{%
\begin{tikzpicture}
\node[station] (base) {#1};

\draw[line join=bevel] (base.100) -- (base.80) -- (base.110) -- (base.70) -- (base.north west) -- (base.north east);
\draw[line join=bevel] (base.100) -- (base.70) (base.110) -- (base.north east);

\draw[line cap=rect] ([yshift=0pt]base.north) [antenna=1];
\end{tikzpicture}
}

\newcommand{\WQU}[1]{%
 \begin{tikzpicture}
\draw[line width=1pt]
        (0,0) -- ++(2cm,0) -- ++(0,-1cm) -- ++(-2cm,0);
\end{tikzpicture}
}

\newcommand{\QU}[1]{%
 \begin{tikzpicture}
\draw[line width=1pt]
        (0,0) -- ++(2cm,0) -- ++(0,-1cm) -- ++(-2cm,0);
        \foreach \Val in {1,...,4}
        \draw ([xshift=-\Val*10pt]2cm,0) -- ++(0,-1cm);
    
\end{tikzpicture}
}

\usepackage{tikz-network}
 
\usepackage{graphicx}

\makeatletter
\def\widebreve{\mathpalette\wide@breve}
\def\wide@breve#1#2{\sbox\z@{$#1#2$}%
     \mathop{\vbox{\m@th\ialign{##\crcr
\kern0.08em\brevefill#1{0.8\wd\z@}\crcr\noalign{\nointerlineskip}%
                    $\hss#1#2\hss$\crcr}}}\limits}
\def\brevefill#1#2{$\m@th\sbox\tw@{$#1($}%
  \hss\resizebox{#2}{\wd\tw@}{\rotatebox[origin=c]{90}{\upshape(}}\hss$}
\makeatletter

 	\definecolor{bluegreen}{rgb}{0.0, 0.87, 0.87}

\begin{document}

	\title{Packet Completion Time Minimization via Joint D2D and  Cellular Communication:  A Unified Network Coding Approach}

\author{Juwendo~Denis,~\IEEEmembership{Member,~IEEE},  
      Hulya~Seferoglu,~\IEEEmembership{Member,~IEEE}
\thanks{Juwendo Denis and   Hulya Seferoglu  are with the Department of  Electrical and Computer Engineering, University of Illinois at Chicago (UIC), Chicago, IL, 60607  (email: jdenis2@uic.edu, hulya@uic.edu )}
\thanks{This work was supported in parts by the Army Research Lab (ARL) under Grant W911NF-1820181 and the National Institute of Standards and Technology (NIST) under Grant 70NANB17H188.
}
}
\maketitle

\begin{abstract}
This paper tackles the problem of transmitting a  common  content  to  a  number  of  cellular  users by means of instantly decodable network coding (IDNC) with the help of  intermittently connected D2D links. Of particular interest are broadcasting real-time applications such as video-on-demand, where common contents may be partially received by cellular users due to packet erasures over  cellular links. Specifically, we investigate the problem of \textit{packet completion time}, defined as the number of transmission slots necessary to deliver a common content to all users.
Drawing on graph theory, we develop an  optimal packet completion time strategy by constructing a two-layer IDNC conflict graph.  The higher-layer graph permits us to determine all feasible packet combinations that can be transmitted over the cellular link, while  the lower-layer graph enables us to find all feasible network coded packets and identify the set of users that can generate and transmit these packets via intermittently connected D2D links. By combining the higher-layer and  the lower-layer IDNC conflict graphs,   we demonstrate that finding the optimal IDNC packets to minimize the packet completion time problem is equivalent to finding the maximum independent set of the two-layer IDNC conflict graph, which is known to be an NP-hard problem.  We design a scheme that invokes the Bron-Kerbosch algorithm  to find the optimal policy.  To circumvent the high computational complexity required to reach the global optimum,  we establish  a polynomial-time solvable low-complexity heuristic  to find  an efficient sub-optimal solution. The effectiveness of our proposed scheme is verified through extensive numerical results which indicate substantial performance improvement in comparison with existing methods.


\end{abstract}


\IEEEpeerreviewmaketitle

\section{Introduction}\label{pap_intro}
In modern communications networks, the number of active and connected devices is expected to increase at an unprecedented rate, reaching 28.5 billion by 2022   \cite{Cisco_2019}. Moreover,  wireless systems are  witnessing an expansion of mobile data traffic on a global scale \cite{Ericsson_2019}.
Therefore,  academic and industry researchers alike have prioritized designing appropriate technologies to prevent this growth in devices and traffic from becoming a major impediment to the performance of 5G and beyond wireless systems. One feasible strategy to address the issue of mobile devices' traffic growth is by employing technologies such as  Device-to-Device (D2D) communications. D2D technology enables   spatially close mobile users to  establish short range communication  without relaying through base stations (BSs). Therefore, D2D presents considerable potential to improve the performance of 5G systems in general \cite{Boccardi_5G_2014}, and dense wireless networks in particular, by facilitating traffic offload.

Performance improvement of wireless systems can also be achieved through the use of network coding (NC), which advocates the transmission of an intelligent combination of packets. NC emerges as a major breakthrough for broadcast wireless channels by enabling efficient packet transmission and achieving judicious resource utilization  through packet coding. Moreover, network coding  facilitates  substantial performance gain in terms of delay reduction and reliable communication \cite{Ahlswede663}. These properties make NC ideal for real-time applications such as mobile video streaming. 

Network coding  has sparked a lot of research interest in the past decade\cite{Tran_4476175,Sorour_5658178,Dong_4578943,Aboutorab_75017,Zayene_8649,Ambadi_6163}. The two most widely  used types of network coding are opportunistic network coding (ONC) \cite{Katt_2005i,Sorour_5658178} and random linear network coding (RLNC) \cite{Ho_1705002, Nisto_5753572,Chiasserini_6600698}. In the latter, the transmitter broadcasts coded packets obtained through a linear combination of all source packets (i.e., all packets available at the transmitter end) by using random coefficients drawn from a finite Galois Field.  Employing RLNC requires that users decode their intended packets only after successfully receiving all independent network coded packets broadcast by the transmitter. Meanwhile, ONC leverages the diversity of the information available at the receivers' end to create coding opportunities to optimize different performance metrics \cite{Douik_689}.  One category of ONC  is instantly decodable network coding (IDNC) \cite{Le_6570827,Sorrour_ACM2015}, which  provides instant packet decoding capability upon successful reception of  the coded packet.  
This eliminates the need for extensive buffer storage,  in contrast to RLNC, and results in lower delays, thereby making IDNC more suitable for time-sensitive applications.  Furthermore,  IDNC only requires a binary XOR ($\oplus$) operation \cite{Katti_12941} to encode and decode transmit and received packets, respectively. This mechanism makes IDNC very appealing  from the perspective of implementation.

\subsection{Related works}
Combining D2D-enabled communications with IDNC  offers broad opportunities to design efficient 5G systems by leveraging the benefits of both techniques. As such, the combination of these technologies has been given a lot of consideration in the literature \cite{Huang_7994922,Huang_8267241,Huang_860,Karim7590137,Douik_8444481, Douik_7176779,Douik_689,Zhan_1595}. The benefits of IDNC for a dual-hop relay-based D2D communications   were explored in \cite{Huang_7994922}  while the performance of an IDNC-assisted dual-hop two-way relay  D2D communications was studied in \cite{Huang_8267241}. 
In \cite{Huang_860}, Huang et al. combined multi-hop D2D  with network coding to improve data transmission efficiency. The authors in
\cite{Karim7590137} investigated the problem of  mean video distortion minimization for an IDNC-based partially connected D2D network. Multi-hop D2D  configuration was considered in \cite{Douik_8444481,Douik_7176779} where  Douik et al. studied the problem of reducing decoding delay.
By considering  the  perspective  of  graph  theory,  the authors in  \cite{Zhan_1595} focused on deriving the coded packets along with the set of transmitting users for a partially connected D2D network.

\subsection{Motivation}
The aforementioned research findings considered only D2D links while neglecting communication on the cellular link. This may lead to substantial performance loss, as argued  in \cite{Seferoglu_6120171}. They demonstrated that concurrent operation on both D2D and cellular links is a viable strategy to enhance  system throughput. Therefore, it is crucial to assess  the performance of network coding-enabled cellular and D2D networking. Particularly, we are interested in evaluating the benefits of using IDNC in the process of packet recovery via both  D2D and cellular  links.   This issue arises in broadcast scenarios where  common contents may be partially received by the users, leading to packet loss generally caused by channel impairments such as multipath fading, pathloss, shadowing and/or severe interference.

Taking into consideration a fully connected D2D network, the authors in \cite{Zhan7947022,Keshtkarjahromi_8259} addressed the problem of packet completion time minimization - the number of transmission slots necessary to recover all missing packets - over joint cellular and D2D systems with IDNC. The authors designed several heuristics to find  feasible packet combinations that can be transmitted by the BS, and to identify IDNC packets as well as the corresponding transmitting users to generate and broadcast the identified codes over D2D links. These works, however, left unaddressed the optimal design of IDNC codes  to minimize the packet completion time using joint cellular  and  D2D communications. Furthermore, \cite{Zhan7947022,Keshtkarjahromi_8259} considered a fully connected D2D network topology, which is unlikely to occur in practice.  

In real-world scenarios, users are sparsely scattered over a large area, and utilize single-hop or multi-hop short range D2D links whenever possible. However,  depending on their locations, some users may be isolated and  unable to use D2D links and can receive their data only via the cellular link. Isolated users are referred to as singleton users. To the best of our knowledge, existing research has neglected singleton users, consequently the schemes proposed in works such as \cite{Huang_7994922,Huang_860,Huang_8267241,Karim7590137,Douik_8444481, Douik_7176779,Douik_689,Zhan_1595,Zhan7947022,Keshtkarjahromi_8259} will fail to work in the presence of isolated users. 

\subsection{Contributions}
Motivated by this observation, we consider a general network topology that takes into consideration the heterogeneity of D2D connections among users. 
We refer to this setup as an \textit{intermittently connected D2D network}. It  includes (i)  single-hop fully connected D2D networks \cite{Zhan7947022,Keshtkarjahromi_8259}, (ii) multi-hop partially connected D2D topology \cite{Huang_860,Karim7590137,Douik_8444481, Douik_7176779,Zhan_1595}, (iii) singleton users, and (iv) disjoint clusters of any combination of the aforementioned configurations. 
In the considered topology, multiple users can transmit their packets simultaneously over the D2D link thanks to disjoint connections between the transmitting users. 
The joint cellular and intermittently connected D2D networking studied in this paper is a general case of the joint cellular and fully connected D2D scenario investigated in \cite{Zhan7947022,Keshtkarjahromi_8259}. To the best of our knowledge, no existing literature has addressed the design of optimal IDNC codes for joint cellular and intermittently connected D2D links, which is the main focus of this paper.

The key contributions of this paper are as follows:
 \begin{enumerate} 
\item  We present a rigorous study to address the problem of broadcasting a common content which, in practice and due to packet erasure channels, may be partially received by some users.    
Particularly, we assess the performance of  efficient packet coding strategies with the help of intermittently connected D2D communications in the process of recovering the missing packets.
\item  Drawing on graph theory, we theoretically characterize  the optimal IDNC policy to minimize the packet completion time  by constructing a  two-layer IDNC conflict graph. The conflict graph takes into consideration all inadmissible packet combinations for the BS and the transmitting D2D users. Specifically, the higher-layer IDNC conflict graph aims to find  all feasible coding opportunities that can be transmitted by the BS over cellular links,  while the lower-layer IDNC conflict graph  determines all feasible packet combinations and identifies the users that can generate and broadcast these packets on the D2D links.

\item By combining the higher-layer IDNC conflit graph  and the lower-layer IDNC conflict graph,   we demonstrate that  solving the problem of packet completion time  is equivalent to finding the maximum independent set of the two-layer IDNC conflict graph. 
\item We proposed a sequential approach, {\tt{OptIDNC}}, based on the Bron-Kerbosch algorithm \cite{bron73} to find the global optimum of the packet completion time problem. This approach, however, entails exponential time complexity, making it less viable for  high-density  networks. 
\item  To circumvent the computational burden inherent to the optimal approach,  we design an efficient polynomial time solvable low-complexity algorithm {\tt{NetCAM-WP}}. The proposed {\tt{NetCAM-WP}}  is capable of finding feasible combination of packets that can be broadcast over both cellular and D2D links, and identifying  transmitting users to generate and broadcast the codes that can be transmitted via D2D links.
\item We also derive an upper-bound on the packet completion time achieved by  the proposed {\tt{NetCAM-WP}}.  
\end{enumerate}

The remainder of this paper is structured as follows:
we introduce the  system model alongside the problem formulation in Section \ref{sec:system-prob}. In 
Section \ref{sec:global_layer}, we characterize the optimal solution and describe  the algorithm to reach the global optimum.  The proposed {\tt{NetCAM-WP}} is introduced in Section \ref{sec:heuristic}, while simulation results are provided in Section \ref{sec:simulation_results}. Finally, we conclude our paper in Section \ref{sec:conclusion}.

\begin{figure}[!]
 \centering
\begin{tikzpicture}[scale=1]
 \begin{scope}
[every node/.append style={draw=gray, left color=white, single arrow},
      transform shape]
\node(E1) [right color=bluegreen,  ellipse, minimum height=5cm,minimum width=8.3cm, opacity=0.2] at (1.1,0){} ; 
 \node(SIN2) [right color=lime,  ellipse, minimum height=1.8cm,minimum width=3.6 cm, rotate=15]at (2.7,-1.3) {};
\node(E4) [right color=pink,  ellipse, minimum height=1.25cm,minimum width=2.8cm, opacity=0.6, rotate=165]at (-1.1,-1.4){}; 
  \end{scope}

 \node(eNB1)[black, scale=0.9] at (1.1,0.6){\BS{BS}};

 \node(A1)[blue, scale=1.4]  at (1.3,-1.4){\UE{UE5}}; 
 \node(A2) [blue, scale=1.4] at (2.5,-1.3){\UE{UE6}};
 \node(A33)[blue, scale=1.4]   at (3.8,-0.7){\UE{UE8}};
 \node(A4)[blue, scale=1.4]   at (3.55,-1.65){\UE{UE7}};
 \node(A6)[red, scale=1.4]   at (4.8,0.5){\UE{UE9}};
  \node(A7)[violet, scale=1.4]   at (-0.25,-1.5){\UE{UE4}};
  \node(AN2)[red, scale=1.4]   at (-2.6,0.5){\UE{UE1}};
  \node(AN)[violet, scale=1.4]   at (-2.2,-1.1){\UE{UE2}};
    \node(AN1)[violet,scale=1.4]  at (-1.2,-1.8){\UE{UE3}};

\draw[black,radiation,decoration={angle=60},scale=0.7] ([xshift=-0.98cm,yshift=-0.25cm]eNB1.north east) -- +(90:0.5);

\draw[blue,radiation,decoration={angle=60},scale=0.25] ([xshift=-1.1cm,yshift=-0.7cm]A1.north east) -- +(90:0.5);
\draw[blue,radiation,decoration={angle=60},scale=0.25] ([xshift=-1.1cm,yshift=-0.7cm]A2.north east) -- +(90:0.5);
\draw[blue,radiation,decoration={angle=60},scale=0.25] ([xshift=-1.1cm,yshift=-0.7cm]A33.north east) -- +(90:0.5);
\draw[blue,radiation,decoration={angle=60},scale=0.25] ([xshift=-1.1cm,yshift=-0.7cm]A4.north east) -- +(90:0.5);
\draw[red,radiation,decoration={angle=60},scale=0.25] ([xshift=-1.1cm,yshift=-0.7cm]A6.north east) -- +(90:0.5);
\draw[violet,radiation,decoration={angle=60},scale=0.25] ([xshift=-1.1cm,yshift=-0.7cm]A7.north east) -- +(90:0.5);
\draw[violet,radiation,decoration={angle=60},scale=0.25] ([xshift=-1.1cm,yshift=-0.7cm]AN.north east) -- +(90:0.5);
\draw[violet,radiation,decoration={angle=60},scale=0.25] ([xshift=-1.1cm,yshift=-0.7cm]AN1.north east) -- +(90:0.5);
\draw[red,radiation,decoration={angle=60},scale=0.25] ([xshift=-1.1cm,yshift=-0.7cm]AN2.north east) -- +(90:0.5);

 \draw [<->,scale=1.5,blue] (A2) -- (A1)node[pos=0.5,sloped, above,scale=0.4] {};
  \draw [ <->,scale=1.5,blue] (A2) -- (A33)node[pos=0.5,sloped, above,scale=0.3] {}; 
     \draw [<->,scale=1.5,blue] (A2) -- (A4)node[pos=0.5,sloped, above,scale=0.3] {};

     \draw [<->,scale=1.5,violet] (AN) -- (AN1)node[pos=0.5,sloped, above,scale=0.2] {};
    \draw [<->,scale=1.5,violet] (AN) -- (A7)node[pos=0.5,sloped, above,scale=0.2] {};
 \draw [<->,scale=1.5,violet] (AN1) -- (A7)node[pos=0.5,sloped, below,scale=0.3] {};

 \draw [dashed, ->](1.1,1.3) to [out=50,in=70](A2){};
  \draw [dashed, ->](1.1,1.3) to [out=50,in=45](A1){};
   \draw [dashed, ->](1.1,1.3) to [out=140,in=25](AN2){}; 
   \draw [dashed, ->](1.1,1.3) to [out=50,in=95](A33){};
    \draw [dashed, ->](1.1,1.3) to [out=50,in=90](A4){}; 
     \draw [dashed, ->](1.1,1.3) to [out=140,in=55](AN){}; 
      \draw [dashed, ->](1.1,1.3) to [out=140,in=90](A7){};  
      \draw [dashed, ->](1.1,1.3) to [out=140,in=80](AN1){};  
             \draw [dashed, ->](1.1,1.3) to [out=50,in=140](A6){};  
  \draw [dashed,-, scale=1.5](-2,1.5) -- (-1.9,1.5)node[pos=1,sloped, right ,scale=0.5] {UE2 to UE4: fully connected D2D};
 \draw [dashed,-, scale=1.5](-2,1.3) -- (-1.9,1.3)node[pos=1,sloped, right ,scale=0.5] {UE5 to UE8: multi-hop D2D};
 \draw [dashed,-, scale=1.5](-2,1.1) -- (-1.9,1.1)node[pos=1,sloped, right ,scale=0.5] {UE1 and UE9: singleton users};

  \draw [<->, scale=1.5](2.6,1.5) -- (2.9,1.5)node[pos=1,sloped, right ,scale=0.5] {D2D link};
  \draw [dashed, ->, scale=1.5](2.6,1.3) -- (2.9,1.3)node[pos=1,sloped, right ,scale=0.5] {Cellular link}; 
\end{tikzpicture}
\caption{Illustration of intermittently connected D2D network where users can recover missing packets via both cellular downlink and D2D links. \label{fig:systemmodel}}
\vspace{-0.6cm}
\end{figure}
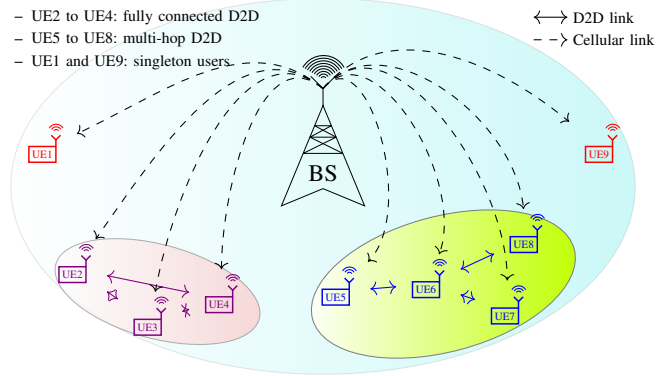

\section{System Setup and Problem Formulation}\label{sec:system-prob}

In this paper, we consider an intermittently connected device-to-device-enabled cellular network topology that  consists of one base station seeking to communicate with a set of $\mathcal{N}=\{1,2, \cdots, N\}$ cellular users. We assume that the users are equipped with D2D interface, as illustrated in Fig.~\ref{fig:systemmodel}. The considered D2D network configuration can be represented by the connection matrix $\bfC  \in \{0,1 \}^{|\mathcal{N}|  \times  |\mathcal{N}| } $ with entries $c_{jk} \in \{0,1 \}$. Specifically,  $c_{jk}$ is set to $1$ if user $j$ is directly connected to user $k$ via a single-hop D2D link, and  $c_{jk}$ is set to $0$, otherwise. We  define the coverage area of each  user as follows:
\begin{defn}\cite{Karim7590137}
The coverage area $\mathcal{Y}_j$ of  user $j$ is defined as the set of neighboring users that are directly connected to it, i.e.,  $\mathcal{Y}_j \triangleq \{ k \in \mathcal{N}| c_{jk}=1 \}$.
\end{defn}
5G and beyond wireless networks are expected to face greater strain on resources  as the number of users continues to grow. Due to the scarcity of radio resources, BSs may rapidly be flooded with increasing and asynchronous requests from a large population of users. One viable way of mitigating this issue  and ensuring efficient spectrum utilization is by leveraging  the broadcast nature of  wireless communications systems \cite{Young-4072749}. Furthermore, for real-time applications such as video-on-demand services including Netflix or YouTube,  a large proportion of traffic is caused by a few popular files that are more likely to be cached at the BS \cite{Golrezaei_6495773}.  

This motivates us to consider  a broadcast scenario where all users are interested in receiving a common popular content from the BS. A content (e.g,  a video or music file) is divided into a set of $\mathcal{M}=\{1, 2, \cdots, M\}$ packets. Moreover, it is assumed that the system operates in two stages.  The BS starts by broadcasting, via cellular links,   $M=|\mathcal{M}|$ packets to all $N=|\mathcal{N}|$  users during the  first stage.
 Unstable variations in the state of the broadcast wireless channel (e.g., fading, shadowing, etc.) may cause packet loss at the users' end. Thus, after the first stage,   the users  inform the BS about their packet reception status via an error-free link \cite{Karim_7864466}. The packet reception status is determined by a feedback matrix  $\bfF \in \{0,1 \}^{M \times  N }$ whose entries  $f_{nm}$  are defined as follows: 
	   \begin{equation}
    \begin{aligned}
    & f_{mn}   \triangleq 
    \left\{
    \begin{array}{ll}
    1  & \mbox{if  packet $m \in \mathcal{H}_n $}  \\
    0 & \mbox{if  packet  $m \in \mathcal{W}_n $} 
    \end{array}
    \right.\\
    \end{aligned}
    \end{equation}
where $ \mathcal{H}_n $, the {\it{Has}} set,  denotes the set of packets that have been successfully received by user $n \in \mathcal{N}$, and $\mathcal{W}_n$,  the {\it{Wants}} set, is the set of packets not received by user $n$  after the first stage. 

In the second stage, referred to as the recovery  phase \cite{Sorrour_ACM2015},  a packet retransmission mechanism is employed to ensure delivery of all missing packets.   Specifically, missing packets are recovered by simultaneous broadcast of efficient packet combinations via both cellular and D2D links. 
In practice,  simultaneous transmission over cellular and D2D links is possible because cellular  and D2D communications use  different frequencies \cite{keller2012microcast, Keshtkarjahromi_8259}.  
During the recovery  phase,  communications either from  the BS to the cellular users or from users to users are assumed to be established via  lossless channels. However, our underlying framework can be easily applied to the lossy-channel scenario. 

\subsection{Instantly decodable network coding}

\begin{Definition}\label{def:network}
A packet  received by a user is instantly decodable if it contains exactly and at most one packet from the  {\it{Wants}} set of the user. 
\end{Definition}
\textit{Example 1}: {\em Benefits of IDNC when cellular links are used.} Consider the following example where a set of three users $\{$UE1, UE2, UE3$\}$ requests four packets $\{ p_1, p_2, p_3, p_4\}$ from the BS. After the first stage, the  {\it{Has}} sets of the users are given by:  $\mathcal{H}_1=\{p_1, p_4 \},\, \mathcal{H}_2=\{ p_1, p_2, p_3\}$ and $\mathcal{H}_3=\{p_2, p_3 \}$ and their  {\it{Wants}} sets by:  $\mathcal{W}_1=\{p_2, p_3 \},\, \mathcal{W}_2=\{ p_4\}$ and $\mathcal{W}_3=\{p_1 , p_4 \}$. Without network coding, the BS needs four transmission slots to recover all missing packets.  By leveraging network coding, only two transmissions slots are required. More precisely, the BS may broadcast coded packet $p_3\oplus p_4$ in the first transmission slot  while transmitting  $p_1\oplus p_2$ in the second one or vice versa, reducing the transmission slots from four to two.

\subsection{Benefits of network coding in the presence of joint D2D/cellular transmission}
The research findings \cite{Huang_7994922,Le_6570827,Sorrour_ACM2015,Huang_860,Huang_8267241,Karim7590137,Douik_8444481, Douik_7176779,Douik_689,Zhan_1595} have exploited the benefits of network coding to improve packet completion time or to reduce decoding delay, using either cellular link or  device-to-device interface separately. If  D2D  and cellular links are used simultaneously, the benefits of IDNC in the process of packet recovery can be further amplified \cite{Zhan7947022,Keshtkarjahromi_8259}. Consider the aforementioned Example 1.  Using only the cellular link, the BS needs two transmission slots  to retransmit all missing packets. Using simultaneously cellular and D2D links, and under the assumption of a fully connected D2D network, UE2 can transmit $p_1 \oplus p_3$ to UE1 and UE3 via D2D link, while the BS broadcasts $p_2\oplus p_4$ to all users over the cellular link. In this case, all packets can be recovered during a single transmission. This is a clear manifestation of how simultaneous use of cellular and D2d links can augment the benefits of network coding. However, the performance improvements of IDNC for  intermittently connected D2D-enabled cellular networks  have yet to be assessed, which is the focus of this paper.

\subsection{Problem Formulation}
Our objective is to establish the optimal design procedure by  determining optimal packet combinations for concurrent transmission over D2D/cellular links, and identifying the optimal scheduling mechanism. Accordingly, we solve the following problem. 

\begin{problem}\label{prob:formulation}
Given an intermittently connected D2D network topology and the feedback matrix  $\bfF$,  
find the optimal IDNC codes to minimize the  packet completion time by exploiting simultaneously the cellular and the D2D links during the recovery phase. 
\end{problem}

\section{Two-Layer IDNC Conflict Graph}\label{sec:global_layer}
We address Problem \ref{prob:formulation}  by leveraging concepts from graph theory. Specifically, we construct a two-layer IDNC conflict  graph that  determines the set of all feasible codes that can be transmitted by the base station, and identifies the set of all users that can create feasible packet combinations and transmits them via D2D links. A similar approach of constructing IDNC graphs has been considered in \cite{Zhan_1595, Karim7590137}, but only for partially connected D2D networks without considering cellular links or isolated users. This indicates that our underlying framework is a general case of existing works \cite{Zhan_1595, Karim7590137}.

\subsection{Two-layer IDNC graph construction}
Before proceeding to describe the two-layer IDNC graph, let us state the following definition.  
\begin{defn}\label{def:inad}({\bf Transmission inadmissibility})
  Transmission inadmissibility occurs when a transmitted coded packet received by a user contains at least two source packets from the {\it{Wants}} set of the user.
\end{defn}
\subsubsection{Higher-layer  IDNC conflict graph}  Built from the perspective of the BS, it intends to create the set of all  feasible packet combinations that can be transmitted over the cellular link. We denote  the higher-layer IDNC graph  as  $\mathcal{G}_1\left(\mathcal{V}_1, \mathcal{E}_1  \right)$, which is an undirected graph constructed with the set of packets from $\mathcal{M}$ yet to be recovered by the users.  The set of vertices $\mathcal{V}_1$ corresponds to the set of all users' missing packets, i.e.,  $\cup_{n \in \mathcal{N} }\mathcal{W}_n$. Each packet $p_l \in \cup_{n \in \mathcal{N} }\mathcal{W}_n$ is associated with one and only one vertex $v_{l}^{(\text{BS})} \in \mathcal{V}_1 $.
The  \textit{conflict} is hidden through the construction of edges between the vertices. Given that a vertex is associated with one and only one packet in $\cup_{n \in \mathcal{N} }\mathcal{W}_n$, whenever the combination of two packets in $\cup_{n \in \mathcal{N} }\mathcal{W}_n$ leads to a transmission inadmissibility,   an edge is established between the two vertices associated with the packets. That is: 
\begin{equation}\nonumber
\forall (v_{n}^{(\text{BS})} , v_{l}^{(\text{BS})} ) \in \mathcal{V}_1, (v_{n}^{(\text{BS})} , v_{l}^{(\text{BS})} ) \in  \mathcal{E}_1  \,\, \text{if} \,\, \exists \, k \in \mathcal{N} | (p_n, p_l) \in \mathcal{W}_k
\end{equation}
\textit{Example 2}: Consider that $4$ users want to receive $3$ packets. Assume that both UE1 and UE3 are connected to UE2 via a single-hop D2D connection, and UE4 is a singleton. The {\it{Has}} and {\it{Wants}} sets of the users are given respectively by $\mathcal{H}_1=\{p_2 \},\, \mathcal{H}_2=\{ p_1,  p_3\}\,$, $\mathcal{H}_3=\{ p_2\},\,\mathcal{H}_4=\{p_2, p_3 \}$ and   $\mathcal{W}_1=\{p_1, p_3 \},\, \mathcal{W}_2=\{ p_2\}, \, \mathcal{W}_3=\{ p_1, p_3\},\,\mathcal{W}_4=\{p_1 \}$. Given that $\cup_{n=1}^4\mathcal{W}_n=\{ p_1,p_2,p_3\}$,  the set $\mathcal{V}_1$ of vertices is given by  $\mathcal{V}_1=\{v_{1}^{\text{BS}}, v_{2}^{\text{BS}}, v_{3}^{\text{BS}} \}$.  There is only one edge in the higher-layer graph, between $v_{1}^{\text{BS}}\text{and}\, v_{3}^{\text{BS}}$, since $p_1 \oplus p_3$ results in transmission inadmissibility.

\subsubsection{Lower-layer IDNC conflict graph}
The graph is constructed from the perspective of all D2D users. Before proceeding to describe it, let us state the following definitions. 
\begin{defn}({\bf Congestion})
Congestion occurs on the D2D links when a user simultaneously receives packets from multiple  users in a single time slot. 
\end{defn}
\begin{defn}\footnote{It should be noted one could technically refer to both Definitions 3 and 4 as collision. However, we  use the terms congestion and conflict to  distinguish between collision that happens in two different scenarios.}({\bf Conflict})
Conflict occurs when two  users that are connected to each other via a single-hop D2D communication initiate  packet transmission simultaneously. 
\end{defn}
The  lower-layer  IDNC conflict  graph is denoted as  $\mathcal{G}_2\left(\mathcal{V}_2, \mathcal{E}_2  \right)$, which is an undirected graph. Its vertices and edges are derived as follow:  
Let $v^{(i)}_{l}$ be associated with the $i$th user ($i \in \mathcal{N}$) and the $l$th packet ($\forall p_l \in \mathcal{M} \cap \mathcal{H}_i$). $v^{(i)}_{l}$  is a vertex of  the graph $\mathcal{G}_2\left(\mathcal{V}_2, \mathcal{E}_2  \right)$ if there exists  a user $k$ in the coverage area of user $i$ that wants packet $p_l$. That is, $\forall i \in \mathcal{N}, v^{(i)}_{l}\in \mathcal{V}_2  \rightarrow \exists (k,l)| \left\{k \in \mathcal{Y}_i \right\} \cap \left\{ p_l  \in \mathcal{H}_i  \cap \mathcal{W}_k \right\}$. Two vertices of the   undirected graph $\mathcal{G}_2\left(\mathcal{V}_2, \mathcal{E}_2  \right)$ are connected by an edge if at least one of the following conditions is satisfied.
\begin{itemize} 
\item[C1] The vertices correspond to two different packets, and the coded combination of the two packets leads to transmission inadmissibility over the D2D links.
\item[C2] The vertices correspond to two different users, which are connected to a third user. The third user  cannot receive packets transmitted simultaneously by the first two users because it will create congestion.
\item[C3]  The vertices correspond to two different users that cannot transmit coded packets  simultaneously because it will create conflict.
\end{itemize}
 Consider the aforementioned \textit{Example 2}. The set of vertices $\mathcal{V}_2 $ includes   $\mathcal{V}_2 =\{v_1^2, v_3^2, v_2^1, v_2^3 \}$. In fact, $\{v_1^2, v_3^2\} \in \mathcal{V}_2$  due to the fact that UE2 can transmit packets $p_1$ and $p_3$ that are needed by UE1. $v_2^1 \in  \mathcal{V}_2$  since UE1 and UE2 are connected and $p_2 \in \mathcal{H}_1  \cap \mathcal{W}_2$. Moreover, $ v_1^2$ and $ v_3^2$ are connected by condition C1.  $v_1^2$ is linked to  $v_2^1$ by an edge due to condition C2. Finally, all pairs of vertices that are connected due to condition C3 are: $(v_2^1, v_1^2)$, $(v_2^1, v_3^2)$,  $(v_2^3, v_1^2)$ and  $(v_2^3, v_3^2)$. 
 \subsubsection{Two-layer IDNC conflict graph} Combines the higher-layer IDNC graph with the lower-layer IDNC graph. The two-layer IDNC conflict graph is  an undirected graph $\mathcal{G}\left(\mathcal{V}, \mathcal{E}  \right)$,  where the vertices and edges are constructed using the set of vertices and edges of both the higher and lower  layer IDNC graphs $\mathcal{G}_1\left(\mathcal{V}_1, \mathcal{E}_1  \right)$ and $\mathcal{G}_2\left(\mathcal{V}_2, \mathcal{E}_2  \right)$, i.e.,  $\mathcal{V}= \mathcal{V}_1 \cup \mathcal{V}_2$. In the combined graph, new edges are created between higher and lower layers to avoid  redundancy,\footnote{Redundancy  leads to a waste of radio resources. Although its occurrence  is not prohibited from the standpoint of  either MAC layer or network layer, for the purpose of our framework's optimality, we assume that it is infeasible.} which occurs  when the same  packet is transmitted simultaneously over both cellular and D2D links. In particular, a vertex from the higher-layer IDNC graph is connected to a vertex  from lower-layer IDNC graph if both vertices are induced by the same packet. In Example 2, the pairs of vertices that are linked by an edge due to redundancy  are given by 
$(v_1^{\text{BS}}, v_1^2)$, $(v_2^{\text{BS}}, v_2^1)$,  $(v_2^{\text{BS}}, v_2^3)$ and  $(v_3^{\text{BS}}, v_3^2)$ because packets $p_1$. In fact, $v_1^{\text{BS}}$ is connected to $ v_1^2$ because they are induced by packet $p_1$. 

\subsection{Characterization of the Optimal solution}
We now proceed to characterize the optimal solution of Problem \ref{prob:formulation}.  We first state the following theorem. 
\begin{Theorem}\label{thm:global_feasibility}
Identifying  users that can create and transmit IDNC packets over D2D links, and finding IDNC packets that  can be broadcast simultaneously over both cellular and D2D links,   is equivalent to finding an independent set of the two-layer IDNC conflict graph $\mathcal{G}\left(\mathcal{V}, \mathcal{E}  \right)$.
\end{Theorem}
\emph{Proof:} The proof of Theorem \ref{thm:global_feasibility} is built  upon the following 2 lemmas. 
\begin{lemma}\label{lem:feas_BS}
Finding an IDNC packet that can be transmitted over the cellular link is equivalent to finding an independent set of the  higher-layer IDNC conflict graph $\mathcal{G}_1\left(\mathcal{V}_1, \mathcal{E}_1  \right)$. 
\end{lemma}
\begin{lemma}\label{lem:feas_D2D}
Finding a set of users that can generate and transmit IDNC packets  as well as their corresponding  feasible IDNC packets is equivalent to finding an independent set of the lower-layer   IDNC conflict graph $\mathcal{G}_2\left(\mathcal{V}_2, \mathcal{E}_2  \right)$.
\end{lemma}
 The proofs of Lemma \ref{lem:feas_BS} and Lemma \ref{lem:feas_D2D} can be found in Appendices \ref{appendix: proofFeasBS} and \ref{appendix:proofFeasD2D}, respectively.

To prove Theorem \ref{thm:global_feasibility}, we start by showing the sufficient condition. Suppose that $\mathcal{N}_2$ is the set of users scheduled to broadcast on the D2D links and $\mathcal{M}_{\text{BS, D2D}}$ is a set of instantly decodable packets that can be transmitted by the BS or any user in  $\mathcal{N}_2$. $\mathcal{M}_{\text{BS, D2D}}$ is the union of $\mathcal{M}_{\text{D2D}}$ and $\mathcal{M}_{\text{BS}}$, where $\mathcal{M}_{\text{BS}}$ is a set of IDNC packets that can be broadcast  from the BS via cellular links, while $\mathcal{M}_{\text{D2D}}$ is a set of instantly decodable packets that can be transmitted from one user to another via D2D links.

With the indices of packets in $\mathcal{M}_{\text{BS}}$, we can generate a set of vertices  $\mathcal{V}_{\text{BS}} \subseteq \mathcal{V}_1\subseteq \mathcal{V}  $. Similarly,  the indices of the users in $\mathcal{N}_2$ together with the indices of their associated IDNC packets enable us to  identify a  set of vertices  $\mathcal{V}_{\text{D2D}} \subseteq \mathcal{V}_2 \subseteq \mathcal{V}  $. Thus, $\mathcal{V}_{\text{BS, D2D}}\triangleq \mathcal{V}_{\text{BS}}\cup \mathcal{V}_{\text{D2D}}$ is the set of vertices associated with feasible codes in $ \mathcal{M}_{\text{BS, D2D}} $. It holds true that $\mathcal{V}_{\text{BS, D2D}} \subseteq  \mathcal{V} $. We show by contradiction that no two vertices in $\mathcal{V}_{\text{BS, D2D}}$ are connected. Suppose that at least two vertices are connected. Given that $\mathcal{M}_{\text{BS}}$ is a set of feasible codes that the BS can transmit via cellular links, we know by Lemma \ref{lem:feas_BS} that no two vertices in $\mathcal{V}_{\text{BS}}$ can be linked by an edge. Connections between at least two vertices in $\mathcal{V}_{\text{BS, D2D}}$ cannot happen in $\mathcal{V}_{\text{BS}}$.
Similarly, using the result of Lemma \ref{lem:feas_D2D}, we can state that the connection will not happen in $\mathcal{V}_{\text{D2D}}$. This means that there  exists at least one vertex in $\mathcal{V}_{\text{BS}}$ that is connected to at least one vertex in $\mathcal{V}_{\text{D2D}}$. Consequently, the vertices correspond to at least one common packet. This leads to  redundancy,  contradicting the fact that $\mathcal{M}_{\text{BS, D2D}}$ is instantly decodable. Therefore, $\mathcal{V}_{\text{BS, D2D}}$ is an independent set. 

Suppose the existence of an independent set  $\widetilde{\mathcal{V}} \subseteq \mathcal{V}$ of the two-layer  graph $\mathcal{G}\left(\mathcal{V}, \mathcal{E}  \right)$. Let $\widetilde{\mathcal{M}}$ be the set of all IDNC packets associated with the vertices in $\widetilde{\mathcal{V}}$. We need to demonstrate that the  codes in $\widetilde{\mathcal{M}}$ are instantly decodable and efficient in the sense that they do not lead to (i) transmission inadmissibility,  (ii) conflict,  (iii) congestion or (iv) redundancy.     We show the instant decodability and efficiency of $\widetilde{\mathcal{M}}$ by considering the following three cases that arise with  $\widetilde{\mathcal{V}}$. 
 \begin{enumerate}
\item  Suppose that $\widetilde{\mathcal{V}} \subseteq \mathcal{V}_1$, i.e., all vertices of  $\widetilde{\mathcal{V}}$ belong to the higher-layer IDNC graph $\mathcal{G}_1\left(\mathcal{V}_1, \mathcal{E}_1  \right)$. Given that $\widetilde{\mathcal{V}} \subseteq \mathcal{V}_1$, it is known from  Lemma  \ref{lem:feas_BS} that $\widetilde{\mathcal{M}}$ is an IDNC packet that can be transmitted over  the cellular link. Given that no  transmission occurs over the D2D link, we  therefore conclude that the codes in $\widetilde{\mathcal{M}}$ are  instantly  decodable. 
\item  Suppose that $\widetilde{\mathcal{V}} \subseteq \mathcal{V}_2$, i.e.,   $\widetilde{\mathcal{V}}$ is in the lower-layer IDNC graph $\mathcal{G}_2\left(\mathcal{V}_2, \mathcal{E}_2  \right)$. Given that $\widetilde{\mathcal{V}} \subseteq \mathcal{V}_2$,  Lemma  \ref{lem:feas_D2D} implies that a set of users that can create and transmit feasible packet combinations can be identified. Furthermore,   $\widetilde{\mathcal{M}}$ determines the corresponding IDNC packets that can be transmitted via D2D links. Since the BS is not broadcasting any packets,  we  conclude that the codes in $\widetilde{\mathcal{M}}$ are instantly decocable and efficient. 
\item Suppose  $\widetilde{\mathcal{V}}$ consists of vertices that are in both $\mathcal{V}_1$ and $\mathcal{V}_2$. We denote $\widetilde{\mathcal{V}}_{\text{BS}}$, the vertices of $\widetilde{\mathcal{V}}$ that are in  $\mathcal{V}_1$, and $\widetilde{\mathcal{V}}_{\text{D2D}}$ those of $\widetilde{\mathcal{V}}$ that are in  $\mathcal{V}_2$. This means that $\widetilde{\mathcal{V}}=\widetilde{\mathcal{V}}_{\text{BS}} \cup \widetilde{\mathcal{V}}_{\text{D2D}}$, and that  $  \widetilde{\mathcal{V}}_{\text{BS}}$ and $ \widetilde{\mathcal{V}}_{\text{D2D}}$ are both independent sets. Following this separation  of $\widetilde{\mathcal{V}}$, we can also break down the set $\widetilde{\mathcal{M}}$ of  IDNC packets associated with the vertices in $\widetilde{\mathcal{V}}$ as  $\widetilde{\mathcal{M}} =\widetilde{\mathcal{M}}_{\text{BS}} \cup \widetilde{\mathcal{M}}_{\text{D2D}}$, where $\widetilde{\mathcal{M}}_{\text{BS}} $ and $ \widetilde{\mathcal{M}}_{\text{D2D}}$ correspond to   $\widetilde{\mathcal{V}}_{\text{BS}}$ and $\widetilde{\mathcal{V}}_{\text{D2D}}$, respectively. Lemma  \ref{lem:feas_BS} indicates that $\widetilde{\mathcal{M}}_{\text{BS}}$ is the set of feasible IDNC packets that the BS can broadcast via cellular links.  Moreover,   Lemma  \ref{lem:feas_D2D} implies that a set of users that can create and transmit the codes in $\widetilde{\mathcal{V}}_{\text{D2D}}$  can be identified. Furthermore,   $\widetilde{\mathcal{V}}_{\text{D2D}}$ determines the corresponding IDNC packets that each scheduled user can transmit via D2D links.  To complete the proof, it remains to show that the codes in $\widetilde{\mathcal{M}}_{\text{BS}} \cup \widetilde{\mathcal{M}}_{\text{D2D}}$ do not lead to redundancy. This is done by contradiction. Suppose that there is at least one pair of codes in $\widetilde{\mathcal{M}}_{\text{BS}} \cup \widetilde{\mathcal{M}}_{\text{D2D}}$ that leads to  redundancy. This means that there is at least one common packet in  both  $\widetilde{\mathcal{M}}_{\text{BS}}$ and $ \widetilde{\mathcal{M}}_{\text{D2D}}$. Therefore, there exists at least one edge between at least one vertex  in $\mathcal{V}_1$ and one vertex in $\mathcal{V}_2$. This leads to a contradiction since  $\widetilde{\mathcal{V}}$ is an independent set. Therefore, the codes in  $\widetilde{\mathcal{M}} = \widetilde{\mathcal{M}}_{\text{BS}} \cup \widetilde{\mathcal{M}}_{\text{D2D}}$ are instantly decodable and efficient. \hfill{$\blacksquare$}
\end{enumerate}

Theorem \ref{thm:global_feasibility} creates a one-to-one mapping between an independent set of the two-layer IDNC conflict graph and combinations of  packets that are instantly decodable and efficient, and can be transmitted via both cellular and D2D links.  This one-to-one mapping  coupled with the following proposition are essential in proving the optimality of our underlying framework.  
\begin{proposition}\label{prop:add_vertglobal}
Let $\overline{\mathcal{V}}\subseteq \mathcal{V}$ be an independent set of the  two-layer IDNC conflict  graph $\mathcal{G}\left(\mathcal{V}, \mathcal{E}  \right)$. Suppose $\overline{\mathcal{N}}$ is the number of users  scheduled to broadcast over the D2D links and $\overline{\mathcal{M}}_{\text{BS, D2D}}$ is the set of IDNC packets associated with $\overline{\mathcal{V}}$.    
Let $ \breve{v}_i^{(j)} \in \{\mathcal{V}\backslash  \overline{\mathcal{V}} \}$ with $j \in (\text{BS}, 1, \cdots, N), \, i \in (1, \cdots M) $ and let $p_{i}$ be the packet associated with vertex $\breve{v}_i^{(j)}$.  $\overline{\mathcal{V}} \cup \{\breve{v}_i^{(j)}\}$ is an independent set of the  two-layer IDNC conflict  graph $\mathcal{G}\left(\mathcal{V}, \mathcal{E}  \right)$ if and only if $\overline{\mathcal{M}}_{\text{BS, D2D}}\oplus p_{i} $ is an efficient IDNC packet. 
Let $ \breve{v}_i^{(j)} \in \{\mathcal{V}\backslash  \overline{\mathcal{V}} \}$ with $j \in (\text{BS}, 1, \cdots, N), \, i \in (1, \cdots M) $ and let $p_{i}$ be the packet associated with vertex $\breve{v}_i^{(j)}$.  $\overline{\mathcal{V}} \cup \{\breve{v}_i^{(j)}\}$ is an independent set of the  global conflict IDNC graph $\mathcal{G}\left(\mathcal{V}, \mathcal{E}  \right)$ if and only if $\overline{\mathcal{M}}_{\text{BS, D2D}}\oplus p_{i} $ is  joint BS/D2D feasible IDNC packet.
\end{proposition}
\emph{Proof:} The proof of Proposition \ref{prop:add_vertglobal} is provided in Appendix \ref{appendix:proofProsiCARDINAL}   \hfill{$\blacksquare$}

\begin{remark}\label{rmk:propoBS} Proposition \ref{prop:add_vertglobal} argues that if a new set created by  adding  a vertex to an independent set is also an independent set, then new IDNC packets can be generated by combining codes that correspond to the original independent set  with  the uncoded packet associated with the new vertex,  and vice versa. Similarly, we can prove that 
if a new independent set is created by removing a vertex from an independent set, then a new IDNC packet can be generated by removing the packet associated with the removed vertex from the  IDNC packet corresponding to the original independent set. This clearly indicates that $|\overline{\mathcal{V}}| = |\overline{\mathcal{M}}_{\text{BS, D2D}}|$. 
\end{remark}
\begin{Corollary}\label{opt_sol}
Finding an optimal strategy at each transmission slot for Problem \ref{prob:formulation} is equivalent to finding the maximum independent set of the two-layer IDNC  conflict  graph $\mathcal{G}\left(\mathcal{V}, \mathcal{E}  \right)$.  
\end{Corollary}
\emph{Proof:} 
 Theorem \ref{thm:global_feasibility} and Proposition \ref{prop:add_vertglobal} indicate  that for each IDNC strategy with a certain number of packets, there exists an independent set of the same size,  and vice versa. The authors in \cite{Le_7568985} proved that an IDNC packet  formed by the maximum number of uncoded packets coincides with the optimal solution. Finally, the IDNC packet composed of the highest number of uncoded packets yields the largest independent set, in terms of the number of vertices. This concludes the proof. \hfill{$\blacksquare$}
 
\begin{Corollary}
Solving Problem \ref{prob:formulation} to the global optimum  is NP-hard. 
\end{Corollary}
\emph{Proof:}  Corollary \ref{opt_sol} established that finding an optimal solution to Problem \ref{prob:formulation} is reduced to the problem of finding  a maximum independent set of $\mathcal{G}\left(\mathcal{V}, \mathcal{E}  \right)$ which is an NP-hard optimization problem. Therefore,  Problem \ref{prob:formulation} is NP-hard. 
 \hfill{$\blacksquare$}

\subsection{Description of the proposed {\tt{OptIDNC}}}
We proceed to describe an algorithm {\tt{OptIDNC}},  to find the global optimum of Problem \ref{prob:formulation}.  The proposed {\tt{OptIDNC}}, summarized in Algorithm \ref{Alg:globaloptimumalgo}, invokes  the Bron-Kerbosch algorithm \cite{bron73} to find a maximum independent set of the two-layer IDNC conflict graph at each transmission slot. 
\begin{breakablealgorithm}
\caption{{\tt{OptIDNC}} that  solves Problem \ref{prob:formulation} to the global optimum}
\label{Alg:globaloptimumalgo}
\begin{algorithmic}[1] 
\STATE \hspace*{\algorithmicindent} \textbf{Input}: $\bfF, \bfC,\, \mathcal{W}_{n}\,  \mathcal{H}_{n}$.
\STATE Initialize $T^{\text{Opt}}= 0 $; 
\WHILE{$\mathbf{1}_{N\times 1}^\top \bfF \mathbf{1}_{N\times 1} \neq N\times M$} 
\STATE  Construct the two-layer IDNC conflict graph $\mathcal{G}\left(\mathcal{V}, \mathcal{E}  \right)$;
\STATE Run the Bron-Kerbosch algorithm  to find the maximum independent set $\mathcal{V}^\star$ of the undirected graph $\mathcal{G}\left(\mathcal{V}, \mathcal{E}  \right)$;
\STATE Use $\mathcal{V}^\star$ to identify the optimal IDNC packets that can be broadcast by the BS;
 \STATE Use $\mathcal{V}^\star$ to find the optimal set of transmitting users along with their corresponding optimal IDNC packets that can be transmitted on D2D links; 
	                \STATE Update  $\bfF$ and $\mathcal{W}_n, \, \forall n,\, \mathcal{H}_n, \, \forall n$
and set $T^{\text{Opt}} \gets T^{\text{Opt}}+1$;
   \ENDWHILE	
  \STATE \hspace*{\algorithmicindent} \textbf{Output}: $T^{\text{Opt}}$. 
\end{algorithmic}
\end{breakablealgorithm}
The proposed {\tt{OptIDNC}} is implemented in a centralized fashion and, at each transmission slot,   invokes the Bron-Kerbosch algorithm \cite{bron73} whose complexity is on the order of  $\mathcal{O}\left(3^{\frac{|\mathcal{V}|}{3}}\right)$. Hence, the complexity of {\tt{OptIDNC}} to solve Problem \ref{prob:formulation} to the global optimum is $\mathcal{O}\left(3^{\frac{|\mathcal{V}|}{3}}\max_{n\in \mathcal{N}}\frac{\mathcal{W}_n}{2}\right)$.  {\tt{OptIDNC}} has exponential time complexity,  which makes it less viable for scenarios with a high-density population of users or large numbers of packets. Accordingly, we propose a low-complexity heuristic to find efficient sub-optimal IDNC codes that can be transmitted via D2D and cellular links  to minimize the packet completion time.

\section{Proposed Heuristic Approach}\label{sec:heuristic}
We proceed to describe our proposed heuristic, \textit{\underline{Net}}work \textit{\underline{C}}oding \textit{\underline{A}}lgorithm  built from the perspective of  \textit{\underline{M}}ost \textit{\underline{W}}anted \textit{\underline{P}}acket ({\tt{NetCAM-WP}}) by the users. At each transmission slot, {\tt{NetCAM-WP}} first determines the most wanted packet $p_{\tt{most}}$ by all  users. Then, it identifies the set of packets that can be combined with $p_{\tt{most}}$ to create a feasible code that the BS can transmit via cellular links. For transmissions over D2D links, {\tt{NetCAM-WP}} identifies the users that can broadcast $p_{\tt{most}}$, and finds a feasible code for each user   without generating either conflict or   congestion. {\tt{NetCAM-WP}} can  be summarized as follows:    

\begin{breakablealgorithm}
\caption{Pseudocode of our proposed {\tt{NetCAM-WP}} }
\label{Alg:proposed_heuristic}
\begin{algorithmic}[1] 
\STATE Initialize $T= 0 $ and identify the set $\mathcal{S}_{\tt{BS}}$ and  $\cup_{n \in \mathcal{N}_{\tt{single}} } \mathcal{W}_{n}$  to be broadcast by the BS; 
\WHILE{$\mathbf{1}_{N\times 1}^\top \bfF \mathbf{1}_{N\times 1} \neq N\times M$}
	            \IF{$  |\mathcal{S}_{\tt{BS}} | \neq 0  $}
	            \STATE  Identify the most wanted packet $p_{\tt{most}}^{\tt{D2D}}$ and  user  $\text{UE}_{p_{\tt{most}}^{\tt{D2D}}}$ that can transmit $p_{\tt{most}}^{\tt{D2D}}$; 
                   \STATE  Find the packets in $\mathcal{H}_{\text{UE}_{p_{\tt{most}}^{\tt{D2D}}}}\backslash \{ p_{\tt{most}}^{\tt{D2D}} \}$ that can be assembled  with $p_{\tt{most}}^{\tt{D2D}}$ to create an IDNC packet; 
		   \STATE Determine  the set $\overline{\mathcal{N}}_{\text{UE}_{p_{\tt{most}}^{\tt{D2D}}}}$ of D2D users that can simultaneously transmit  with $\text{UE}_{p_{\tt{most}}^{\tt{D2D}}}$  over the D2D links; 
\STATE Determine the IDNC packets to be transmitted by each user in $\overline{\mathcal{N}}_{\text{UE}_{p_{\tt{most}}^{\tt{D2D}}}}$;       
	            \STATE The BS broadcasts $ p_i \in \mathcal{S}_{\tt{BS}} $ to all users over the cellular link and  update  $\mathcal{S}_{\tt{BS}}  \gets \mathcal{S}_{\tt{BS}} \backslash p_i$;
\ELSE \IF{$\mathcal{S}_{\tt{BS}} = \emptyset$ AND $|\cup_{n \in \mathcal{N}_{\tt{single}} } \mathcal{W}_{n}| \neq 0$}
\STATE Choose any packet $p_{\tt{single}} \in \cup_{n \in \mathcal{N}_{\tt{single}} } \mathcal{W}_{n}$; 
\STATE Find the set $\mathcal{S}_{\text{singleton}}$ of packets that can be combined with $p_{\tt{single}}$ to create an IDNC packet; 
\STATE The BS transmits $ \mathcal{S}_{\tt{singleton}} \oplus p_{\tt{single}}$ over cellular links;
\STATE Repeat Step 4  to Step 7 to determine the set of D2D users and their associated IDNC packet that can be transmitted over be the D2D link;
\ENDIF
	            \ELSE
	            \STATE Find the most wanted packet $p_{\tt{most}}$ by the users; 
\IF{$ \tilde{n} >1$}
 \STATE Identify the set $\mathcal{S}_{\tt{most}}^i$  that can be combined with $p_{\tt{most}}^i$ to create the IDNC packet $\mathcal{S}_{\tt{most}}^i \oplus p_{\tt{most}}^i, \, \forall i=1, \cdots \tilde{n}$;
 \STATE Select the combination with the maximum number of receivers;  
  \ELSE
   \STATE Find the set of packets  to assemble  with $p_{\tt{most}}$, with the goal of creating an IDNC packet that can be broadcast by the BS;
\ENDIF           
   \STATE Repeat Step 4  to Step 7 to identify the set of D2D users along with their associated IDNC packet that can be transmitted over the D2D links;
	            \ENDIF
	                \STATE Update  $\bfF$ and set $T \gets T+1$;
   \ENDWHILE	
\end{algorithmic}
\end{breakablealgorithm}
The detailed description of {\tt{NetCAM-WP}} is provided as follows: {\tt{NetCAM-WP}} starts by identifying the set $\widehat{\mathcal{S}}_{\tt{BS}}$ of packets that can be transmitted only by the BS. $\widehat{\mathcal{S}}_{\tt{BS}}$  includes (i) $\mathcal{S}_{\tt{BS}}$, defined as the set of packets that belong  to the {\it{Wants}} set of all the users, i.e., 
$\mathcal{S}_{\tt{BS}}\triangleq\left\{ p_i, \forall i \in \mathcal{M}| p_i \in \mathcal{W}_n, \forall n \in \mathcal{N}\right\}$
and (ii) the {\it Wants} set of all singleton users, i.e., $\cup_{n \in \mathcal{N}_{\tt{single}} } \mathcal{W}_{n}$, where $\mathcal{N}_{\tt{single}}$ represents the set of all singleton users. 

For the cellular link, the BS  removes one packet from $\mathcal{S}_{\tt{BS}}$ at each transmission slot until $\mathcal{S}_{\tt{BS}}$ becomes an empty set.  
Once $\mathcal{S}_{\tt{BS}}$ is empty,  {\tt{NetCAM-WP}} randomly selects and removes one packet, $p_{\tt{single}}$,  from  $\cup_{n \in \mathcal{N}_{\tt{single}} } \mathcal{W}_{n}$. Then,  {\tt{NetCAM-WP}} identifies the set $\mathcal{S}_{\tt{single}}  $ of packets  from $\mathcal{M}\backslash \{ p_{\tt{single}} \} $ that can be combined with $p_{\tt{single}}$ to form an instantly decodable packet $ \mathcal{S}_{\tt{single}} \oplus p_{\tt{single}}$ to be transmitted  by the BS over the cellular link. The process is repeated until the set $\widehat{\mathcal{S}}_{\tt{BS}} $ is empty. 

After transmitting all packets in $\widehat{\mathcal{S}}_{\tt{BS}}$,    {\tt{NetCAM-WP}} finds at each transmission slot  
$p_{\tt{most}}=\arg \max_{p_i, \forall i \in 1,\cdots, |\mathcal{M}|}\cap_{n\in\mathcal{N}} \mathcal{W}_n$.  
Then, {\tt{NetCAM-WP}}  identifies the set $\mathcal{S}_{\tt{most}} \in \{p_1, p_2, \cdots, p_{M} \}\backslash p_{\tt{most}}$ that can be assembled with $p_{\tt{most}}$ to create an instantly decodable packet $\mathcal{S}_{\tt{most}} \oplus p_{\tt{most}}$ beneficial to all users. In some cases, there might be several packets that are wanted by most users. Whenever this occurs, we denote  $\tilde{n}$ the number of packets most wanted by the cellular users and $p_{\tt{most}}^1, \cdots, p_{\tt{most}}^{\tilde{n}}$, the packets. For each packet $p_{\tt{most}}^i \in \{ p_{\tt{most}}^1, \cdots, p_{\tt{most}}^{\tilde{n}}\}$, {\tt{NetCAM-WP}}  identifies the set $\mathcal{S}_{\tt{most}}^i$ of packets that can be combined with  $p_{\tt{most}}^i$ to create an instantly decodable packet. Then,  {\tt{NetCAM-WP}} selects from  $\mathcal{S}_{\tt{most}}^1 \oplus p_{\tt{most}}^1, \cdots, \mathcal{S}_{\tt{most}}^{\tilde{n}} \oplus p_{\tt{most}}^{\tilde{n}} $ the combination with the largest number of packets. If $n_1$ packet combinations have the largest number of packets, {\tt{NetCAM-WP}} selects the code with the maximum number of receivers from the $n_1$ packet combinations. 

For the D2D links, {\tt{NetCAM-WP}}  identifies at each transmission slot the packet that most users require by  excluding the set $\mathcal{P}_{\text{BS}}$ of packets selected to be transmitted over the cellular link. This can be written as $
p_{\tt{most}}^{\tt{D2D}}=\arg \max_{p_i, \forall i \in 1,\cdots, |\mathcal{M}|}\{\cap_{n\in\mathcal{N}} \mathcal{W}_n\}\backslash \mathcal{P}_{\text{BS}}
$. When several packets are most wanted by the users, one of them is chosen randomly. Then,  {\tt{NetCAM-WP}} selects the user, $\text{UE}_{p_{\tt{most}}^{\tt{D2D}}}$, than can transmit $p_{\tt{most}}^{\tt{D2D}}$ as follows: (i) $p_{\tt{most}}^{\tt{D2D}}$   is in the  {\it{Has}} set of the user and (ii) in the event where $p_{\tt{most}}^{\tt{D2D}}$  belong to {\it{Has}} set of several users, the user with more demand for $p_{\tt{most}}^{\tt{D2D}}$ in its coverage area is selected. 

Once $\text{UE}_{p_{\tt{most}}^{\tt{D2D}}}$ is selected, {\tt{NetCAM-WP}} determines the set $\mathcal{S}_{p_{\tt{most}}^{\tt{D2D}}} \in \mathcal{H}_{\text{UE}_{p_{\tt{most}}^{\tt{D2D}}}}\backslash \{ p_{\tt{most}}^{\tt{D2D}} \}$ of packets  that can be assembled  with $p_{\tt{most}}^{\tt{D2D}}$ to create an IDNC packet beneficial to all users that are in the coverage area of $\text{UE}_{p_{\tt{most}}^{\tt{D2D}}}$. 
Then, {\tt{NetCAM-WP}} identifies  the set $\overline{\mathcal{N}}_{\text{UE}_{p_{\tt{most}}^{\tt{D2D}}}}$ of  users that can concurrently transmit  with $\text{UE}_{p_{\tt{most}}^{\tt{D2D}}}$  over the D2D links without causing  congestion. For each user in $\overline{\mathcal{N}}_{\text{UE}_{p_{\tt{most}}^{\tt{D2D}}}}$, {\tt{NetCAM-WP}} finds the IDNC packet that it can  transmit.        

We now proceed to investigate the performance of {\tt{NetCAM-WP}} by analytically deriving a bound on the  number of transmission slots required by {\tt{NetCAM-WP}} to recover all missing packets. This is stated in the following lemma. 
  \begin{lemma}\label{lemmaNo_tbound}
    The optimal packet completion time $T^{\star}$ achieved by  {\tt{NetCAM-WP}} is bounded by 
\begin{equation}\label{eq:comple:hshs}
\begin{aligned}
|\mathcal{S}_{\tt{BS}}| \leq T^{\star}~&\!\!\! \leq |\mathcal{S}_{\tt{BS}}\cup_{n \in \mathcal{N}_{\tt{single}} } \mathcal{W}_{n}|  + \left \lceil \frac{| \mathcal{W}_{n_{\max}^\star} \backslash \{\mathcal{W}_{n_{\max}^\star} \cap \{\mathcal{S}_{\tt{BS}} \cup_{n \in \mathcal{N}_{\tt{single}} } \mathcal{W}_{n}\} \} |}{2}    \right \rceil 
\end{aligned}
\end{equation}
where $n_{\max}^\star= \arg \max_{n \in \mathcal{N}} |\mathcal{W}_n|$. 
    \end{lemma} 
\emph{Proof:} The proof can be found in Appendix  \ref{appendix:proofFirs_Bound}. \hfill{$\blacksquare$}

\begin{figure}[!]
\begin{minipage}[b]{0.45\linewidth}
\centering
\includegraphics[width=\textwidth]{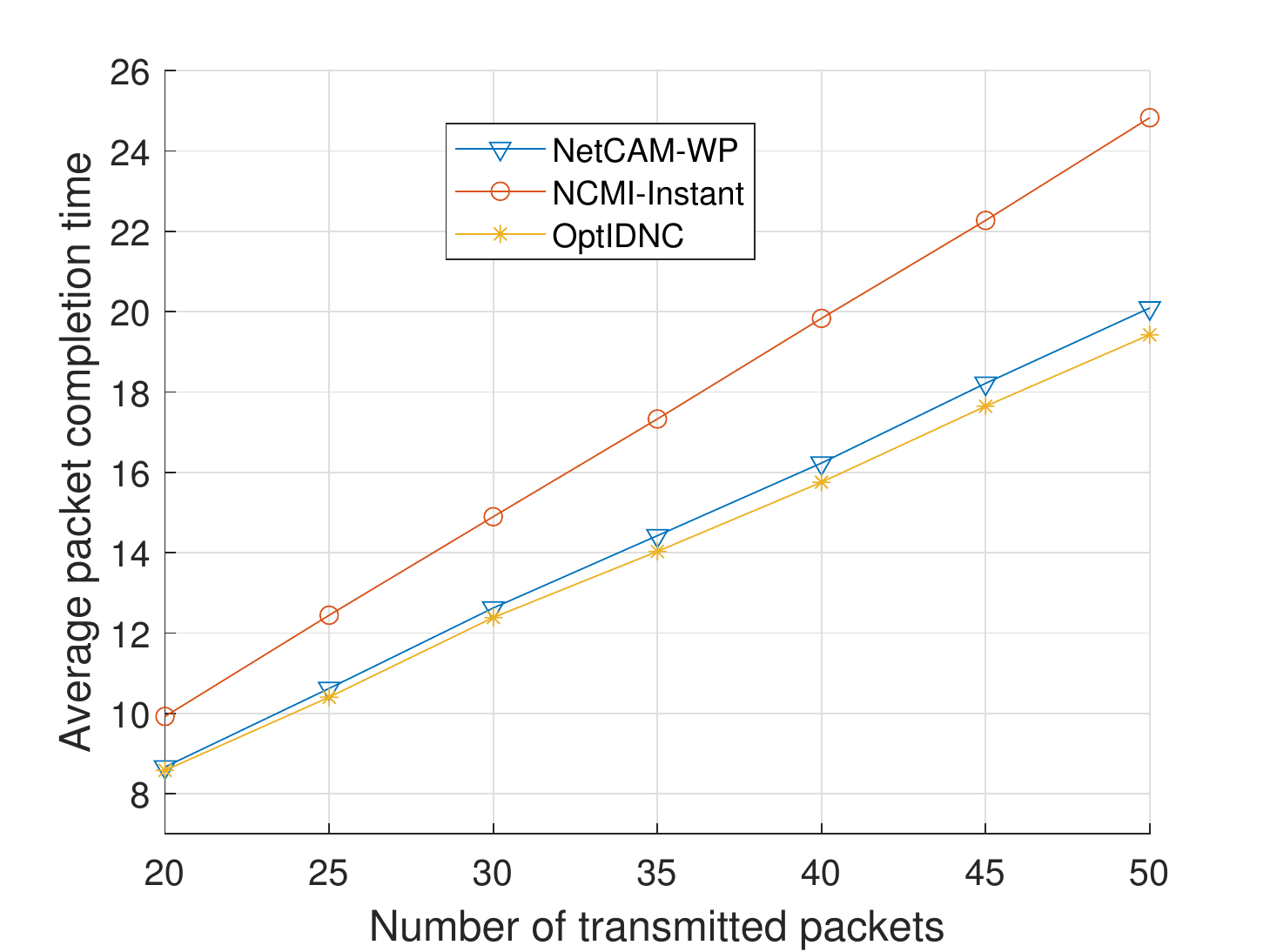}
\caption{Average packet completion time for $N=10$}
\label{fig:fullyconnected}
\end{minipage}
\hspace{0.5cm}
\begin{minipage}[b]{0.45\linewidth}
\centering
\includegraphics[width=\textwidth]{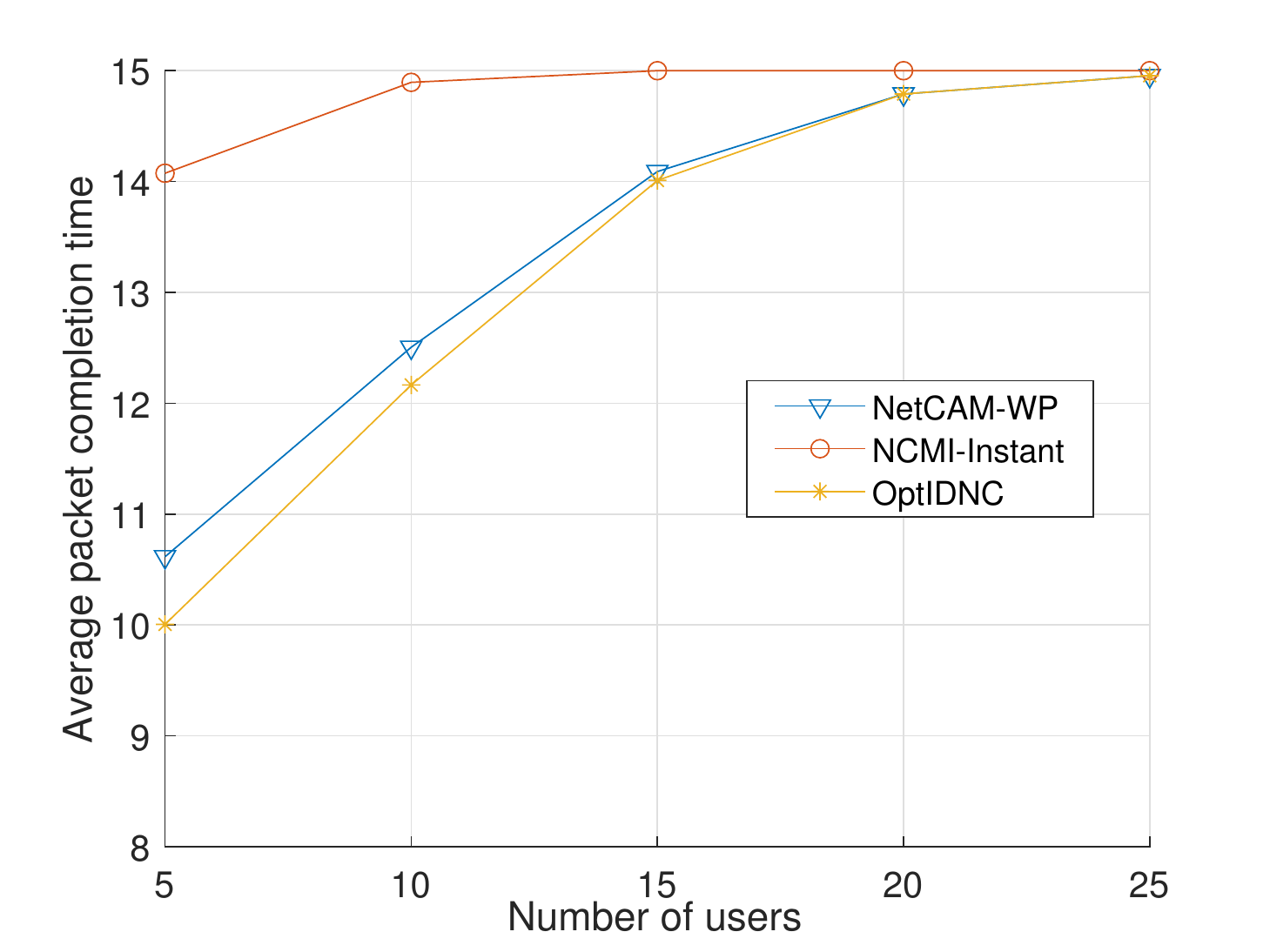}
\caption{Average packet completion time versus number of users for $|\mathcal{M}|=30$}
\label{fig:fullyConnected_Complet_NumberUsers}
\end{minipage}
\end{figure}

\section{Simulation Results}\label{sec:simulation_results}
In this section, numerical results are presented to evaluate the performance of the proposed {\tt{NetCAM-WP}} described in Algorithm \ref{Alg:proposed_heuristic}. All results are obtained using Monte Carlo simulations by averaging over 500 trials on either the connection matrix or the feedback matrix. Unless stated otherwise, the connection matrix $\bfC$ is generated randomly according to a uniform distribution without prior knowledge about either the number of singleton users or the number of disjoint clusters of D2D configurations.  

\subsection{Fully connected D2D network}
We consider a single-hop fully connected D2D network topology which,  unless stated otherwise,  consists of 10 users. As a benchmark, we provide the performance of the {\it network coding for multiple interface (NCMI)-instant} ({\tt{NCMI-Instant}}) algorithm \cite[Algorithm 1]{Keshtkarjahromi_8259}. Fig. \ref{fig:fullyconnected} depicts the performance of {\tt{OptIDNC}}, {\tt{NetCAM-WP}} and {\tt{NCMI-Instant}} in terms of average packet completion time versus the number of packets.   Fig. \ref{fig:fullyconnected} shows that {\tt{NetCAM-WP}} outperforms the existing {\tt{NCMI-Instant}} scheme. Moreover,  as can be seen from  Fig. \ref{fig:fullyconnected},  the gap between the performance of {\tt{OptIDNC}} and the one of the proposed {\tt{NetCAM-WP}} is negligible. There is a gain varying from 0.96\% to 2.78\% between the performance of both approaches.

Fig. \ref{fig:fullyConnected_Complet_NumberUsers} portrays the performance in terms of average packet completion time versus the number of users for a fixed number of transmitted packets $|\mathcal{M}=30|$. It can be inferred from Fig. \ref{fig:fullyConnected_Complet_NumberUsers} that the proposed {\tt{NetCAM-WP}} does not only achieve substantial performance improvement compared to the existing {\tt{{\tt{NCMI-Instant}}}}, its performance approaches the one of {\tt{OptIDNC}}. Fig. \ref{fig:fullyconnected}   and Fig. \ref{fig:fullyConnected_Complet_NumberUsers} lead us to conclude that the proposed {\tt{NetCAM-WP}} can  yield  near-optimal solutions at least for low-scale fully connected D2D networks.

\subsection{Intermittently connected D2D network}
\begin{figure}[!]
\begin{minipage}[b]{0.45\linewidth}
\centering
\includegraphics[width=\textwidth]{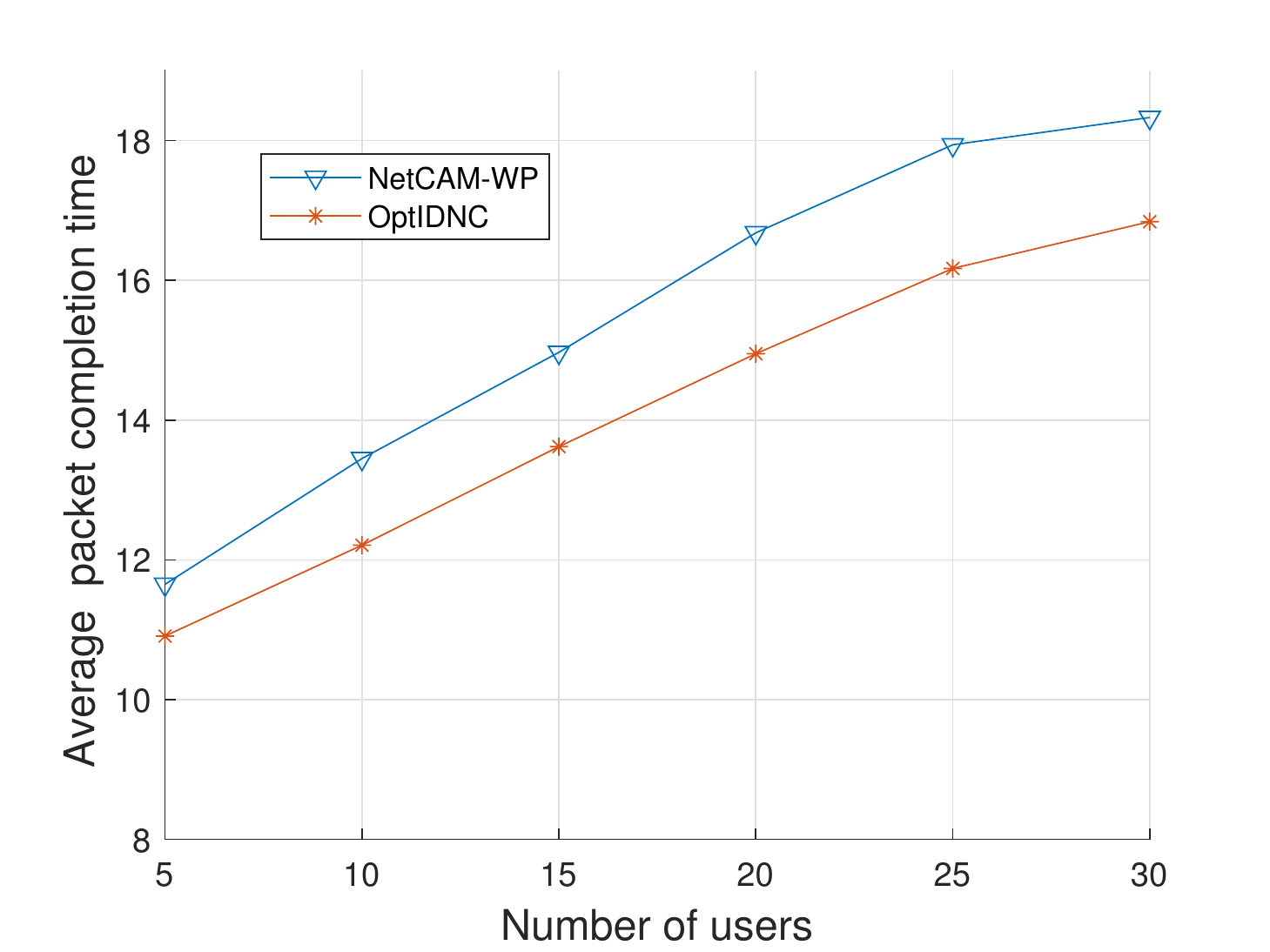}
\caption{Average packet completion time versus number of D2D nodes for $|\mathcal{M}|=25$}
\label{fig:PartiallyConnect_Complet_NumberUsers}
\end{minipage}
\hspace{0.5cm}
\begin{minipage}[b]{0.45\linewidth}
\centering
\includegraphics[width=\textwidth]{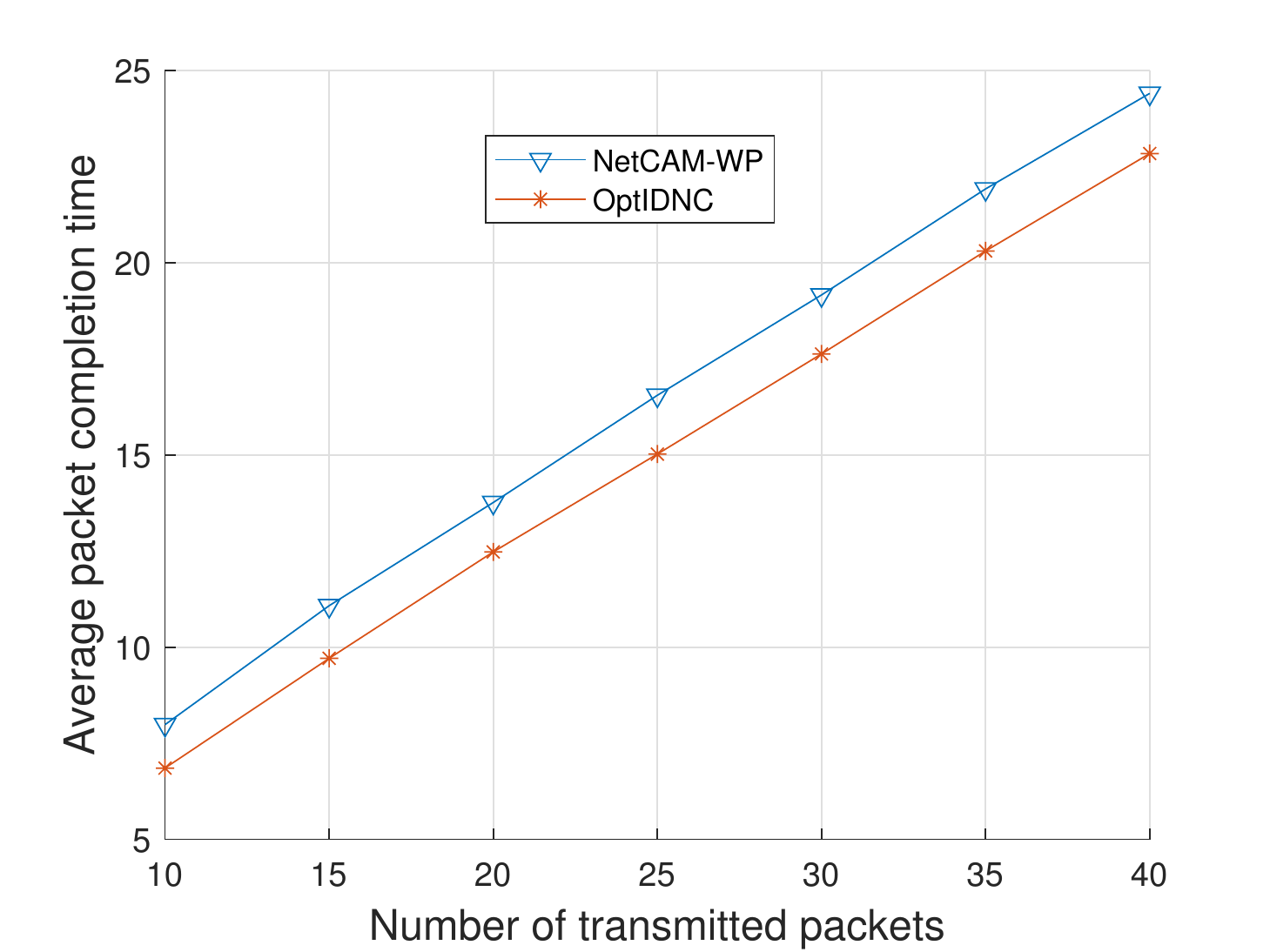}
\caption{Average completion time versus number of transmitted packets for $|\mathcal{N}|=20$}
\label{fig:PartiallyConnect_Complet_PACKETS}
\end{minipage}
\end{figure}

We evaluate the performance of our proposed schemes for an intermittently connected D2D configuration. Fig. \ref{fig:PartiallyConnect_Complet_NumberUsers} depicts the performance of {\tt{NetCAM-WP}} and {\tt{OptIDNC}}  from the perspective of packet completion time versus number of users. One observation that can be drawn from Fig. \ref{fig:PartiallyConnect_Complet_NumberUsers} is that the packet completion time increases as the number of active users grows. 

 Fig. \ref{fig:PartiallyConnect_Complet_PACKETS} portrays the performance of our proposed schemes in terms of evolution of the packet completion time versus the number of transmitted packets. For this scenario, a network topology that consists of  $|\mathcal{N}=20|$ users is studied. In contrast to the  fully connected D2D configuration, we can see from Fig. \ref{fig:PartiallyConnect_Complet_NumberUsers}  and Fig. \ref{fig:PartiallyConnect_Complet_PACKETS} that the  gap between the performance of {\tt{NetCAM-WP}} and the performance of {\tt{OptIDNC}} becomes wider. One plausible explanation is  that, under an intermittently connected D2D setup, the configuration of the users might impact the number of transmitted packets at each transmission slot, increasing the packet completion time. 
 
\section{Conclusion}\label{sec:conclusion}
We considered  a broadcast scenario where a BS seeks to transmit a common content to all users. Part of the transmitted content may be missing at the users' end due to packet loss caused by channel impairments. We argued that the missing packets can be recovered by employing network coding and  intermittent D2D connections between the users. Specifically, we addressed the problem of packet completion time using  graph theory. Accordingly,  we identified feasible solutions through the construction of a two-layer IDNC conflict graph, which enabled us to find IDNC packets to be broadcast over cellular and D2D links along with the set of users that can generate and transmit these codes over the D2D links. Furthermore, we proved that the global optimum is obtained by finding the maximum independent set of the two-layer graph. We designed a sequential mechanism,  {\tt{OptIDNC}},  to reach the optimum solution. We also developed a low-complexity time solvable heuristic, {\tt{NetCAM-WP}},  which was demonstrated through simulation results to achieve a near-optimal solution and to outperform existing approaches for fully connected D2D topology. The simulations also indicated an increase in the gap between the proposed heuristic and the  {\tt{OptIDNC}} under intermittently connected D2D configurations. A plausible future research direction can be to explore mechanisms to further improve the performance of {\tt{NetCAM-WP}} without degrading its complexity.     

 \appendices {\setcounter{equation}{0}
\renewcommand{\theequation}{A.\arabic{equation}} }    
\section{Proof of Lemma \ref{lem:feas_BS} }\label{appendix: proofFeasBS}

We start by showing the necessity of the condition. Suppose that there exists an independent set $\widetilde{\mathcal{V}}_1 \subseteq \mathcal{V}_1  $  of the  higher-layer IDNC graph $\mathcal{G}_1\left(\mathcal{V}_1, \mathcal{E}_1  \right)$. Let $v_{\pi(1)}^{(\text{BS})} , v_{\pi(2)}^{(\text{BS})} , \cdots, v_{\pi(|\widetilde{\mathcal{V}}_1|)}^{(\text{BS})} $ be the elements of $\widetilde{\mathcal{V}}_1$, where $\pi: \{1, 2,\cdots,  |\widetilde{\mathcal{V}}_1|\} \rightarrow \{1,2, \cdots, |\mathcal{M} | \}$ is one-to-one and onto mapping that  associates the index of a vertex  $\widetilde{\mathcal{V}}_1$ to its corresponding packet index in $\mathcal{M}$. 
Consider combinations of all uncoded packets associated with the vertices in $\widetilde{\mathcal{V}}_1$, i.e.,  $p=p_{\pi(1)}\oplus p_{\pi(2)}\oplus\cdots \oplus p_ {\pi(|\widetilde{\mathcal{V}}_1|)} $. We need to show that $p$ is an IDNC packet. This is done by contradiction. Suppose that  $p$ leads to transmission inadmissibility.  Hence,  $p$ is not instantly decocable to at least one user  $u_i \in \mathcal{N}$.  Consequently, at least two packets from $\{p_{\pi(1)}, p_{\pi(2)},\cdots , p_ {\pi(|\widetilde{\mathcal{V}}_1|)} \}$ are also in $\mathcal{W}_{u_i}$, the {\it{Wants}} set of  $u_i $. Therefore, there exists  edges among all vertices associated with the packets (at least two) in  $\{p_{\pi(1)}, p_{\pi(2)},\cdots , p_ {\pi(|\widetilde{\mathcal{V}}_1|)} \} \cap \mathcal{W}_{u_i}$. This contradicts the fact that $\widetilde{\mathcal{V}}_1$ is an independent set.

We proceed to prove the sufficiency of the condition. Suppose that there exists  an IDNC packet, $\tilde{p}$, that the BS can broadcast via cellular links.  Suppose that $\tilde{p}$ is created by combining   $M_1 \leq |\mathcal{M}|$ uncoded packets. Hence,  $\tilde{p}$ can be written as  $\tilde{p}= p_{\check{\pi}(1)}\oplus\cdots \oplus p_{\check{\pi}(M_1)}$ where  $\check{\pi}: \{1, \cdots, M_1\} \rightarrow \{1, \cdots, |\mathcal{M} | \}$. Denote $\widehat{\mathcal{V}}_1$ as the  set  all vertices associated with the packets  $\tilde{p}$. Hence, $\widehat{\mathcal{V}}_1$ is given by  $\widehat{\mathcal{V}}_1=\{v_{\check{\pi}(1)}^{(\text{BS})} , \cdots v_{\check{\pi}(M_1)}^{(\text{BS})}  \}$. It is straightforward to see that  $\widehat{\mathcal{V}}_1 \subseteq  \mathcal{V}_1$. Given that $\tilde{p}$ is instantly decodable  to all users,  no two packets from  the set  $\{ p_{\check{\pi}(1)}, \cdots , p_{\check{\pi}(M_1)}\}$ are simultaneously in the  {\it{Wants}} set of any user. Consequently, there cannot be edges that connect the  vertices  in $\widehat{\mathcal{V}}_1$. Therefore, $\widehat{\mathcal{V}}_1 \subseteq  \mathcal{V}_1$ is an independent set. 
 \hfill{$\blacksquare$}

\section{Proof of Lemma \ref{lem:feas_D2D}}\label{appendix:proofFeasD2D}

To prove Lemma \ref{lem:feas_D2D}, we start by establishing the necessity of the condition.   Suppose that there exists an independent set  $\widetilde{\mathcal{V}}_2 \subseteq \mathcal{V}_2  $  of the graph $\mathcal{G}_2\left(\mathcal{V}_2, \mathcal{E}_2  \right)$. Let $v_{\tilde{\pi}(1)}^{(\hat{\pi}(1))},v_{\tilde{\pi}(2)}^{(\hat{\pi}(2))}, \cdots, v_{\tilde{\pi}(|\mathcal{V}_2|)}^{(\hat{\pi}(|\mathcal{V}_2|))}  $ be the elements of $\widetilde{\mathcal{V}}_2$, where $\tilde{\pi}: \{1, 2,\cdots,  |\widetilde{\mathcal{V}}_2|\} \rightarrow \{1,2, \cdots, |\mathcal{M} | \}$ is a  mapping that associates the subscript of a vertex to a packet in  $\mathcal{M}$, and $\hat{\pi}: \{1, 2,\cdots,  |\widetilde{\mathcal{V}}_2|\} \rightarrow \{1,2, \cdots, |\mathcal{N} | \}$ is a mapping that  matches the superscript of a vertex to a user in $\mathcal{N}$.  It should be   demonstrated that  transmissions resulting from the combinations of any number of vertices of  $\widetilde{\mathcal{V}}_2$ do not lead to (i) transmission inadmissibility,  (ii) conflict or (iii)  congestion. 
Without loss of generality, consider the combination of two vertices, $(v_{\tilde{\pi}(i)}^{(\hat{\pi}(k))},v_{\tilde{\pi}(j)}^{(\hat{\pi}(l))}) \in \widetilde{\mathcal{V}}_2$. We have either one of the following two scenarios: (a) $\tilde{\pi}(i)=\tilde{\pi}(j)$ or (b) $\tilde{\pi}(i)\neq \tilde{\pi}(j)$. 

If $\tilde{\pi}(i)=\tilde{\pi}(j)$, then it can be inferred from the superscripts and subscripts of the two vertices  $v_{\tilde{\pi}(i)}^{(\hat{\pi}(k))}, v_{\tilde{\pi}(i)}^{(\hat{\pi}(l))}$,  that user $\tilde{\pi}(i)$ is broadcasting coded packet $p_{\hat{\pi}(k)} \oplus p_{\hat{\pi}(l)}$. We prove by contradiction that  $p_{\hat{\pi}(k)} \oplus p_{\hat{\pi}(l)}$ does not lead to transmission inadmissibility. Suppose that  $p_{\hat{\pi}(k)} \oplus p_{\hat{\pi}(l)}$ leads to transmission inadmissibility.  Hence, there exists at least one  user in the coverage of  user  $\tilde{\pi}(i)$ that needs to recover both packets $p_{\hat{\pi}(k)} $ and $p_{\hat{\pi}(l)}$. Therefore,  $v_{\tilde{\pi}(i)}^{(\hat{\pi}(k))}$ is connected to $v_{\tilde{\pi}(i)}^{(\hat{\pi}(l))}$, which contradicts the fact that  $\widetilde{\mathcal{V}}_2 $ is an independent set. 

If $\tilde{\pi}(i)\neq \tilde{\pi}(j)$, then one can deduce  from the two vertices  $v_{\tilde{\pi}(i)}^{(\hat{\pi}(k))}, v_{\tilde{\pi}(i)}^{(\hat{\pi}(l))}$,   that user $\tilde{\pi}(i)$  transmits packet $p_{\hat{\pi}(k)}$ while user $\tilde{\pi}(j)$ sends packet $ p_{\hat{\pi}(l)}$. We establish by contradiction that neither conflict nor congestion occurs. 
\begin{enumerate}
\item Suppose that  conflict occurs. Hence,  user $\tilde{\pi}(i)$ is directly connected to user $\tilde{\pi}(j)$.  Therefore, there exists an edge between  $v_{\tilde{\pi}(i)}^{(\hat{\pi}(k))}$ and  $ v_{\tilde{\pi}(i)}^{(\hat{\pi}(l))}$.  This leads to a contradiction since $\widetilde{\mathcal{V}}_2 $ is an independent set. 
\item Suppose that congestion occurs. Hence,  there is at least one user in $\mathcal{Y}_{\tilde{\pi}(i)}\cap \mathcal{Y}_{ \tilde{\pi}(j)}$, i.e., simultaneously in the coverage area of user $\tilde{\pi}(i)$ and user $\tilde{\pi}(j)$.  Therefore, there is an edge that connects $v_{\tilde{\pi}(i)}^{(\hat{\pi}(k))}$ and  $ v_{\tilde{\pi}(i)}^{(\hat{\pi}(l))}$. This is in contradiction with the fact that  $\widetilde{\mathcal{V}}_2 $ is an independent set. 
\end{enumerate}
  To prove the sufficiency of the condition, let $\mathcal{N}_1$ be the number of users scheduled to broadcast on  D2D links. Denote $\hat{p}_n$ the IDNC packet that user $n \in \mathcal{N}_1$ can transmit. 
  Using $n$ as superscript and the indices of all the uncoded packets that create $\hat{p}_n$ as subscript, we can generate vertices for the set $\widehat{\mathcal{V}}_2 $. By repeating this process for all $n \in \mathcal{N}_1$, we can create the set $\widehat{\mathcal{V}}_2 $ of vertices.  Given that $\hat{p}_1, \hat{p}_2, \cdots, \hat{p}_{|\mathcal{N}_1|}$ are feasible codes that can be transmitted  simultaneously over D2D links, they do not cause  conflict,   congestion or transmission inadmissibility.   Hence, there cannot be edges that  connect any  two vertices in $\widehat{\mathcal{V}}_2 $. Therefore, $\widehat{\mathcal{V}}_2 $  is an independent set.   \hfill{$\blacksquare$}

\section{Proof of Proposition \ref{prop:add_vertglobal} }\label{appendix:proofProsiCARDINAL}

Suppose that $\overline{\mathcal{V}} \cup \{\breve{v}_i^{(j)}\}$ is an independent set of the  two-layer IDNC conflict graph $\mathcal{G}\left(\mathcal{V}, \mathcal{E}  \right)$. By Theorem \ref{thm:global_feasibility},
there exists a set  $\breve{p}_{\text{BS, D2D}}$ of coded packets associated with $\overline{\mathcal{V}} \cup \{\breve{v}_i^{(j)}\}$ that is instantly decodable and efficient in the sense that the set $\breve{p}_{\text{BS, D2D}}$ does not lead to  (i) transmission inadmissibility,  (ii)  conflict,   (iii) congestion, or (iv)  redundancy. We demonstrate by contradiction that $\breve{p}_{\text{BS, D2D}} = \overline{\mathcal{M}}_{\text{BS, D2D}}\oplus  p_{i}$  by considering the following three scenarios. 
\begin{enumerate}
\item Suppose that $\overline{\mathcal{V}} \cup \{\breve{v}_i^{(j)}\} \subseteq \mathcal{V}_1$. All vertices are in the higher-layer IDNC graph.  $\breve{p}_{\text{BS, D2D}} \neq  \overline{\mathcal{M}}_{\text{BS, D2D}}\oplus  p_{i}$ means that  $ \overline{\mathcal{M}}_{\text{BS, D2D}}\oplus  p_{i}$ leads to transmission inadmissibility.  Hence, there exists a user that requires $p_{i}$ in addition to a given packet in $\overline{\mathcal{M}}_{\text{BS, D2D}}$. Therefore, there is an edge that connects $\{\breve{v}_i^{(\text{BS})} \}$ to  a vertex in $\overline{\mathcal{V}}$. This leads to a contradiction.  
\item  Suppose that $\overline{\mathcal{V}} \cup \{\breve{v}_i^{(j)}\} \subseteq \mathcal{V}_2$. Given that  $\overline{\mathcal{N}}$ is the number of users scheduled to broadcast via D2D links,   $\overline{\mathcal{M}}_{\text{BS, D2D}}$ consists of the feasible codes that can be transmitted by the $\overline{\mathcal{N}}$ users. Let   $\overline{p}_j, \, \forall j \in (1,\cdots, \overline{\mathcal{N}})$ be the IDNC packet associated with user $j$.  If $ \overline{\mathcal{M}}_{\text{BS, D2D}}\oplus  p_{i}$ is not feasible, then at least one of the combination $\overline{p}_k \oplus  p_{i}, \, k \in (1,\cdots,  \overline{\mathcal{N}}) $ is not a feasible coded packet. Without loss of generality, assume that  $ \overline{p}_{ \overline{\mathcal{N}}} \oplus  p_{i}$ is infeasible. Hence, we have the following two cases. 
\begin{enumerate}
\item The  user with index $j$ coincides with the user labeled as $ \overline{\mathcal{N}}$. From this  user' perspective, there exists at least one user in its coverage area that requires $p_{i} $ in addition to one packet $p_{\tilde{n}} \in \overline{p}_{ \overline{\mathcal{N}}}$ leading to transmission inadmissibility. Therefore,  there is an edge that connects $v_{\tilde{n}}^{( \overline{\mathcal{N}})} \in \overline{\mathcal{V}}$ to $\breve{v}_i^{( \overline{\mathcal{N}})}$. This contradicts the fact that  $ \overline{\mathcal{V}} \cup \{\breve{v}_i^{( \overline{\mathcal{N}})}\}$ is an independent set. 
\item  The user labeled with index $j$ is different from the one  labeled as  index $ \overline{\mathcal{N}}$. Hence, either congestion  or  conflict occurs. Therefore, there is an edge between the vertex $\breve{v}_i^{(j)}$  and all vertices, $\tilde{\mathcal{V}}_{ \overline{\mathcal{N}}} \subseteq \overline{\mathcal{V}}$ which corresponds to the packets in  $ \overline{p}_{ \overline{\mathcal{N}}} $. This contradicts the fact that  $\overline{\mathcal{V}} \cup \{\breve{v}_i^{(j)}\}$ is an independent set. 
\end{enumerate}
\item Suppose that $\overline{\mathcal{V}}$ consists of vertices in  both $\mathcal{V}_1$ and $\mathcal{V}_2$. Denote  $\overline{\mathcal{V}}_{\text{BS}}$, the vertices of $\overline{\mathcal{V}}$ that are in  $\mathcal{V}_1$  and  $\overline{\mathcal{V}}_{\text{D2D}}$, the set  vertices of $\overline{\mathcal{V}}$ that are in  $\mathcal{V}_2$. We can see  that  $\overline{\mathcal{V}}=\overline{\mathcal{V}}_{\text{BS}} \cup \overline{\mathcal{V}}_{\text{D2D}}$ and both sets $ \overline{\mathcal{V}}_{\text{BS}}$ and $\overline{\mathcal{V}}_{\text{D2D}}$ are independent sets. Vertex $\{\breve{v}_i^{(j)}\}$ can be either in the lower-layer  or higher-layer graph. Suppose it is in the lower-layer IDNC conflict graph. The  packet combination $\overline{\mathcal{M}}_{\text{BS, D2D}}\oplus  p_{i}$ is infeasible, which means hat it causes  redundancy. Hence, there is at least one edge that connects a vertex in $ \overline{\mathcal{V}}_{\text{D2D}} \cup \{\breve{v}_i^{(j)}\}$   with a vertex in $\overline{\mathcal{V}}_{\text{BS}}$. This is in contradiction with the fact that $\overline{\mathcal{V}}_{\text{BS}}\cup \overline{\mathcal{V}}_{\text{D2D}} \cup \{\breve{v}_i^{(j)}\} $ is an independent set. 
\end{enumerate}

We start to establish  the sufficiency of the condition by assuming that $\overline{\mathcal{M}}_{\text{BS, D2D}}\oplus  p_{i}$ is instantly decodable and efficient. Suppose that   $\overline{\mathcal{V}} \cup \{\breve{v}_i^{(j)}\}$ is not an  independent set of the  two-layer IDNC conflict graph  $\mathcal{G}\left(\mathcal{V}, \mathcal{E}  \right)$. Hence,  there exists either an  edge that connects  $\{\breve{v}_i^{(j)}\}$ to at least one vertex in $ \mathcal{V}_1$ or an edge that links  $\{\breve{v}_i^{(j)}\}$  to at  least one vertex in $\mathcal{V}_2$. This leads to a contradiction since  $\overline{\mathcal{M}}_{\text{BS, D2D}}\oplus  p_{i}$ instantly decodable and efficient.
\hfill{$\blacksquare$}

\section{Proof of Lemma \ref{lemmaNo_tbound}}\label{appendix:proofFirs_Bound}
 Denote $n_{\max}^\star= \arg \max_{n \in \mathcal{N}} |\mathcal{W}_n|$. To prove Lemma \ref{lemmaNo_tbound}, we consider the following two cases:
\begin{enumerate}
\item Suppose that $n_{\max}^\star \in  \mathcal{N}_{\tt{single}}$. In that case, all packets in $\mathcal{W}_{n_{\max}^\star}$ are transmitted within  the transmission of all  packets in $\mathcal{S}_{\tt{BS}} \cup_{n \in \mathcal{N}_{\tt{single}} } \mathcal{W}_{n}$. The BS broadcasts all packets $\mathcal{S}_{\tt{BS}} \cup_{n \in \mathcal{N}_{\tt{single}} } \mathcal{W}_{n}$ via cellular links and the packets in $\mathcal{M}\backslash\{ \mathcal{S}_{\tt{BS}} \cup_{n \in \mathcal{N}_{\tt{single}} } \mathcal{W}_{n}\}$ are transmitted via D2D links. Therefore, the packet completion time is upper-bounded by $|\mathcal{S}_{\tt{BS}} \cup_{n \in \mathcal{N}_{\tt{single}} } \mathcal{W}_{n}|$. 
\item Suppose that  $n_{\max}^\star \in  \mathcal{N}\backslash \mathcal{N}_{\tt{single}}$.We need to consider two cases: 
 \begin{enumerate}
\item  If $|\mathcal{W}_{n_{\max}^\star}| \leq |\mathcal{S}_{\tt{BS}} \cup_{n \in \mathcal{N}_{\tt{single}} } \mathcal{W}_{n}|$, then the number of transmission slots required to deliver all missing packets is bounded by $|\mathcal{S}_{\tt{BS}} \cup_{n \in \mathcal{N}_{\tt{single}} } \mathcal{W}_{n}|$.
\item If $|\mathcal{W}_{n_{\max}^\star}| \geq |\mathcal{S}_{\tt{BS}} \cup_{n \in \mathcal{N}_{\tt{single}} } \mathcal{W}_{n}|$,  then it holds true that $\mathcal{S}_{\tt{BS}} \subseteq \mathcal{W}_{n_{\max}^\star}$ and $|\cup_{n \in \mathcal{N}_{\tt{single}} } \mathcal{W}_{n}| \leq \mathcal{W}_{n_{\max}^\star}$. Either one the following three scenarios can occur.
\begin{enumerate}
\item  $\cup_{n \in \mathcal{N}_{\tt{single}} } \mathcal{W}_{n} \subseteq \mathcal{S}_{\tt{BS}}$. This case may lead to two possibilities:
\begin{enumerate}
\item $| \mathcal{W}_{n_{\max}^\star} \backslash \mathcal{S}_{\tt{BS}} | \leq  |  \mathcal{S}_{\tt{BS}}|$. While the BS is broadcasting the packets from $\mathcal{S}_{\tt{BS}}$ via the cellular link, user  $n_{\max}^\star$ can recover its missing packets, i.e., $\mathcal{W}_{n_{\max}^\star} \backslash \mathcal{S}_{\tt{BS}}$ via D2D  transmissions. Hence, the packet completion time is bounded by   $|  \mathcal{S}_{\tt{BS}}|$ which is equal to $|\mathcal{S}_{\tt{BS}} \cup_{n \in \mathcal{N}_{\tt{single}} } \mathcal{W}_{n}|$. 
\item  $| \mathcal{W}_{n_{\max}^\star} \backslash \mathcal{S}_{\tt{BS}} | \geq  |\mathcal{S}_{\tt{BS}} |$. After receiving the first $|\mathcal{S}_{\tt{BS}}|$ packets,  there will be at most $| \mathcal{W}_{n_{\max}^\star} \backslash \mathcal{S}_{\tt{BS}} | $  packets missing from the \textit{Wants} set of user $n_{\max}^\star$ which the user can recover during $\left\lceil\frac{| \mathcal{W}_{n_{\max}^\star} \backslash \mathcal{S}_{\tt{BS}} |}{2}\right \rceil$ transmission slots. Therefore, the packet completion time is given by $ |  \mathcal{S}_{\tt{BS}}| +\left\lceil\frac{| \mathcal{W}_{n_{\max}^\star} \backslash \mathcal{S}_{\tt{BS}} |}{2}\right \rceil$.
\end{enumerate}
\item $ \mathcal{S}_{\tt{BS}} \subseteq \cup_{n \in \mathcal{N}_{\tt{single}} } \mathcal{W}_{n}$. This leads to two possibilities:
\begin{enumerate}
\item $| \mathcal{W}_{n_{\max}^\star} \backslash \mathcal{S}_{\tt{BS}} | \leq  | \cup_{n \in \mathcal{N}_{\tt{single}} } \mathcal{W}_{n}|$. The BS transmits the packets from $ \cup_{n \in \mathcal{N}_{\tt{single}} } \mathcal{W}_{n}$ via  cellular links while user  $n_{\max}^\star$ receives its remaining missing packets via D2D transmissions.  Hence, the packet completion time is bounded by $| \cup_{n \in \mathcal{N}_{\tt{single}} } \mathcal{W}_{n}|$. 
\item  $| \mathcal{W}_{n_{\max}^\star} \backslash \mathcal{S}_{\tt{BS}} | \geq  |\cup_{n \in \mathcal{N}_{\tt{single}} } \mathcal{W}_{n}|$. In this case, user    $n_{\max}^\star$  needs to recover at most  $| \mathcal{W}_{n_{\max}^\star} \backslash \{\mathcal{W}_{n_{\max}^\star} \cap \{\cup_{n \in \mathcal{N}_{\tt{single}} } \mathcal{W}_{n}\} \} | $ packets after the first   $|\cup_{n \in \mathcal{N}_{\tt{single}} } \mathcal{W}_{n}|$ transmission slots. This can be achieved during  $\left \lceil \frac{| \mathcal{W}_{n_{\max}^\star} \backslash \{\mathcal{W}_{n_{\max}^\star} \cap \{\cup_{n \in \mathcal{N}_{\tt{single}} } \mathcal{W}_{n}\} \} |}{2}    \right \rceil$ transmission slots. Hence, the packet completion time  is bounded  by $|\cup_{n \in \mathcal{N}_{\tt{single}} } \mathcal{W}_{n}| + \left \lceil \frac{| \mathcal{W}_{n_{\max}^\star} \backslash \{\mathcal{W}_{n_{\max}^\star} \cap \{\cup_{n \in \mathcal{N}_{\tt{single}} } \mathcal{W}_{n}\} \} |}{2}    \right \rceil$.
\end{enumerate}
\item $ \cup_{n \in \mathcal{N}_{\tt{single}} } \mathcal{W}_{n} \nsubseteq \mathcal{S}_{\tt{BS}}$ and  $ \mathcal{S}_{\tt{BS}} \nsubseteq \cup_{n \in \mathcal{N}_{\tt{single}} } \mathcal{W}_{n}$. This leads to two possibilities: 
\begin{enumerate}
\item $| \mathcal{W}_{n_{\max}^\star} \backslash \mathcal{S}_{\tt{BS}} | \leq  |\mathcal{S}_{\tt{BS}}  \cup_{n \in \mathcal{N}_{\tt{single}} } \mathcal{W}_{n}|$. We can use  similar reasoning as in part (i-A) to show that the packet completion time is   bounded by $ |\mathcal{S}_{\tt{BS}}  \cup_{n \in \mathcal{N}_{\tt{single}} } \mathcal{W}_{n}|$. 
\item  $| \mathcal{W}_{n_{\max}^\star} \backslash \mathcal{S}_{\tt{BS}} | \geq  |\mathcal{S}_{\tt{BS}} \cup_{n \in \mathcal{N}_{\tt{single}} } \mathcal{W}_{n}|$. User    $n_{\max}^\star$  needs to recover at most  $| \mathcal{W}_{n_{\max}^\star} \backslash \{\mathcal{W}_{n_{\max}^\star} \cap \{\mathcal{S}_{\tt{BS}}\cup_{n \in \mathcal{N}_{\tt{single}} } \mathcal{W}_{n}\} \} | $ packets after the first   $|\mathcal{S}_{\tt{BS}}\cup_{n \in \mathcal{N}_{\tt{single}} } \mathcal{W}_{n}|$ transmission slots. 
Therefore, the packet completion time is bounded  by $  \left \lceil \frac{| \mathcal{W}_{n_{\max}^\star} \backslash \{\mathcal{W}_{n_{\max}^\star} \cap \{\mathcal{S}_{\tt{BS}} \cup_{n \in \mathcal{N}_{\tt{single}} } \mathcal{W}_{n}\} \} |}{2}    \right \rceil + |\mathcal{S}_{\tt{BS}}\cup_{n \in \mathcal{N}_{\tt{single}} } \mathcal{W}_{n}| $. 
\end{enumerate}
\end{enumerate} 
\end{enumerate}
\end{enumerate}
This concludes the proof. 
\hfill{$\blacksquare$}

\bibliographystyle{IEEEtran}
\bibliography{references}

\end{document}